\documentclass[10pt,journal,comsoc]{IEEEtran}
\IEEEoverridecommandlockouts
\usepackage[T1]{fontenc}% optional T1 font encoding
\usepackage{amssymb}
\usepackage{longtable}
\usepackage{amsmath}
\usepackage{amsthm}
\usepackage{amsfonts}
\usepackage{amsthm}
\usepackage{amsmath}
\usepackage{amsfonts}
\usepackage{amssymb}
\usepackage{graphicx}
\usepackage{hyperref}
\usepackage{algorithmic}
\usepackage{graphicx}
\usepackage{textcomp}
\usepackage{xcolor}
\usepackage{balance}
\usepackage{mathtools}
\usepackage{dsfont}
\usepackage{verbatim}
\usepackage{titlesec}
\titleformat{\subparagraph}[runin]{\normalfont\normalsize\bfseries}{\thesubparagraph}{1em}{}[.]
\titlespacing*{\subparagraph}{0pt}{3.25ex plus 1ex minus .2ex}{1em}
\usepackage{float}
\usepackage{eucal}
\usepackage{balance}
\usepackage{bm}
\usepackage{booktabs}
\usepackage{amssymb}
\usepackage{amsmath}
\usepackage{amsthm}
\usepackage{amsfonts}
\usepackage{relsize}
\usepackage{hyperref}
\usepackage{bbold}
\usepackage{mathtools}
\usepackage{dsfont}
\usepackage{verbatim}
\usepackage{enumitem}
\usepackage{siunitx}
\usepackage{float}
\usepackage{multirow}
\usepackage{fancyhdr}

\titlespacing{\subsection}{3pt}{1.5ex plus 1ex minus .2ex}{0.75ex plus .2ex}

\newtheorem{theorem}{Theorem}%[section]
\newtheorem{cor}{Corollary}
\newtheorem{lemma}{Lemma}

\newtheorem{remark}{Remark}
\newtheorem{definition}{Definition}

\newcommand{\centredComma}{\mathpunct{\raisebox{0.5ex}{,}}}

\begin{document}
%\title{Achievable Spectral Efficiency Over Terrestrial Coherent FSO Links Using Adaptive Transmissions}

%%choice no. 1
\title{Maximum Spectral Efficiency of Adaptive Coherent Terrestrial FSO Links}

%%choice no. 2
%\title{Achievable Spectral Efficiency Limits of Adaptive Coherent Terrestrial FSO Links}
%\title{On the Achievable Spectral Efficiency of Terrestrial Coherent Free-Space
%Optical Links Using Adaptive Transmission}

%choice no. 2
%\title{On the Achievable Spectral Efficiency of Terrestrial Coherent FSO Links Using Adaptive Transmissions}

%% choice no. 2
%\title{On Achievable Spectral Efficiency of Terrestrial\\
%Coherent FSO Links With Adaptive Transmission}

%\title{Maximum Achievable Spectral Efficiency With Adaptive Transmissions Over Terrestrial Coherent FSO Links}

%\title{Achievable Spectral Efficiency With Adaptive Transmissions Over Terrestrial Coherent FSO Channels}
%\title{Achievable Spectral Efficiency With Adaptive FSO Transmissions Over Terrestrial Coherent FSO Links}
%\title{New Insights Into Achievable Spectral Efficiency of Adaptive Coherent FSO Links}

%\title{Novel Insights Into Achievable Spectral Efficiency With Adaptive Coherent FSO Transmissions Over Atmospheric Turbulence Channels}

%\title{New Insights Into Achievable Spectral Efficiency With Adaptive Coherent FSO Transmissions}

%\title{Average Spectral Efficiency With Adaptive Transmissions over Terrestrial Coherent FSO Channels}

\author{
\IEEEauthorblockN{Himani~Verma, Kamal~Singh, and Ranjan~K.~Mallik,~\IEEEmembership{Fellow,~IEEE}}
\vspace{-0.67cm}
\thanks{{Himani~Verma and Kamal~Singh are with the Department of Electrical Engineering, Shiv Nadar Institution of Eminence, Greater Noida, Delhi NCR, 201314, India (e-mail: \tt{hv790@snu.edu.in}; \tt{kamal.singh@snu.edu.in}).}

{Ranjan~K.~Mallik is with the Department of Electrical Engineering, Indian Institute of Technology, Delhi, Haus Khas, New Delhi 110016, India (email: \tt{rkmallik@ee.iitd.ernet.in}).}
}
}%

\maketitle

\thispagestyle{plain}
\pagestyle{plain}

\begin{abstract}
Terrestrial free-space optical (FSO) communication systems, while designed to operate on large unlicensed optical bandwidths, are fundamentally power-constrained due to stringent eye-safety regulations. Moreover, channel fluctuations inherent to terrestrial FSO links further reduce the received optical power. Consequently, the achievable signal-to-noise ratio (SNR) per hertz could become limited---especially for future terrestrial FSO systems based on coherent communications. This necessitates the development of efficient and adaptive communication strategies at both the optical transmitter and receiver. However, a comprehensive assessment of adaptive coherent terrestrial FSO systems remains largely unexplored in the existing literature. This paper investigates terrestrial FSO communication links employing adaptive coherent transmission and synchronous heterodyne detection-based reception operating in the shot-noise-limited regime. Specifically, we propose a novel exact closed-form expression for the average spectral efficiency limit of coherent FSO communication systems with optimal adaptive signaling over gamma-gamma turbulence channels with pointing errors. More importantly, we provide a detailed assessment of the impact of turbulence and pointing error impairments on the coherent FSO system performance, revealing several novel and counterintuitive insights. Extensive numerical results help elucidate the intricacies of analyzing these terrestrial FSO systems and clarify a few misconceptions alluded to in recent related literature.
\end{abstract}


\vspace{-0.5em}
\section{Introduction}\label{sec:intro}
Terrestrial free-space optical (FSO) communication has established itself as a complementary---and in some deployment scenarios, a viable alternative---solution to conventional RF (radio-frequency) and fiber-optic communication technologies for delivering ultra-high-capacity fronthaul and backhaul connectivity. This growth is driven by several key advantages of FSO systems, including vast unlicensed bandwidth, low-cost transceiver implementation, rapid deployment capability, and inherent physical-layer security \cite{arnon2012advanced}. Consequently, the scope of FSO-based communication technologies has steadily broadened and now encompasses a diverse range of application scenarios, including wireless ad hoc networks for tactical and emergency response situations \cite{arnon2012advanced}, terrestrial last-mile broadband networks \cite{stanton2005}, deep-space communications \cite{cornwell2014laser}, inter-satellite communication links \cite{lutzer2010}, and mission-critical civil and military operations \cite{haan2013free}.

A fundamental design consideration in FSO systems concerns how the optical field is exploited for information transfer. Accordingly, two principal transmission and detection paradigms can be identified, with coherent approaches---most commonly realized via heterodyne reception---gradually gaining prominence alongside conventional noncoherent schemes:\\ 
\noindent\emph{\textbf{(i) Intensity modulation with direct detection (IM-DD)}}: In this scheme, information is encoded solely in the intensity of the transmitted optical field, thereby utilizing only \emph{one degree of freedom}. At the receiver, the optical field is detected by direct (square-law) photodetection without phase recovery. IM-DD systems are widely adopted in commercial FSO deployments due to their simple low-cost transceiver architectures; however, the limited sensitivity of IM-DD receivers restricts their applicability to short-range terrestrial FSO links \cite{heatley1998optical}, \cite{barry90}.\\
\noindent\emph{\textbf{(ii) Coherent (field) modulation with synchronous heterodyne detection (HD)}}: Coherent FSO communication systems exploit both the amplitude and phase of the transmitted optical field, thereby leveraging \emph{two degrees of freedom} to convey information. In an HD receiver, a local optical oscillator---typically frequency-offset from the received carrier---is mixed with the incoming optical field prior to photodetection. This process enhances receiver sensitivity and preserves the full complex envelope of the received optical field. Moreover, coherent HD receivers enable highly selective narrowband RF filtering, in contrast to the inherently wideband optical filtering in IM-DD FSO systems. Collectively, these capabilities position coherent communication techniques as strong candidates for next-generation high-capacity terrestrial FSO links \cite{barry90}.

\textcolor{black}{Recent advances in coherent optical transceiver techniques tailored to terrestrial FSO links have renewed research interest, with field trials demonstrating per-channel data rates beyond 100~Gbps \cite{Parca2013}, 400~Gbps \cite{Guiomar2021}, and 800~Gbps \cite{Guiomar2022} (see also references therein). While these demonstrations clearly validate the feasibility of coherent FSO communications in terrestrial environments, they also highlight the nontrivial signal processing burden and practical design constraints imposed by the turbulent optical medium, as noted by Guiomar et al. in  \cite{Guiomar2024dsp}.} Chief among these challenges are channel impairments intrinsic to terrestrial FSO links: turbulence-induced fading and pointing (misalignment) errors. Turbulence-induced refractive-index inhomogeneities in the atmosphere cause random intensity (irradiance) fluctuations, which are typically characterized by statistical channel models such as the lognormal, gamma-gamma (GG), and Málaga distributions, depending on the prevailing turbulence regime \cite{andrews2005laser}. Misalignment induced by structural motion, mechanical disturbances, and strong winds impose additional power penalties, thereby degrading the performance of coherent terrestrial FSO systems. In practice, the severity of pointing errors depends strongly on the acquisition, tracking, and pointing (ATP) subsystem, which mitigates beam misalignment but cannot fully eliminate residual jitter \cite{andrews2005laser}.

\textcolor{black}{Beyond turbulence-induced and pointing-induced irradiance fluctuations, optical power emission in terrestrial FSO links is constrained by stringent eye and skin safety regulations \cite{IEC60825_2014}. When coupled with the vast optical bandwidths available for ultra-high-speed transmission \cite{bloom2003understanding}, stringent power limits can render the available power per hertz a dominant performance bottleneck. While such low-SNR operation is rarely encountered today, recent demonstrations of extreme-wideband wavelength-division multiplexed (WDM) coherent FSO systems indicate that allocating the power-constrained optical budget across many optical carriers can plausibly lead to reduced per-channel SNR \!\cite{marco2024main}, \cite{dochhan2019}, particularly under variable atmospheric attenuation (e.g., haze, fog, or pollution). Accordingly, the low-SNR regime is likely to gain practical relevance in future ultra-wideband WDM terrestrial FSO links. Moreover, irradiance fluctuations further degrade the received optical power~\cite{andrews2005laser}; however, this effect can be alleviated through adaptive transmission strategies. For example, threshold-based transmission---where the transmitter is activated only when the instantaneous channel quality exceeds a predefined level---can be especially effective under stringent power constraints. Importantly, terrestrial FSO links exhibit extremely large coherence bandwidths (hundreds of gigahertz or more) \cite{hong1976,Lu2012}, and millisecond-scale coherence times \cite{mostafa2012}, thereby enabling a quasi-static, frequency-flat block-fading channel. Such a block-fading channel structure facilitates adaptive coherent transmission, as demonstrated in recent field experiments~\cite{Guiomar2021,Guiomar2022,marco2024main,dochhan2019}.}

%These channel impairments pose significant challenges for maintaining reliable high-throughput links in coherent FSO systems where phase integrity is critical. Drawing parallels to adaptive techniques in mobile wireless communications~\cite{goldsmithbook2005},~\cite{tsebook2005}, rate and power adaptation emerges as a promising strategy to enhance both robustness and spectral efficiency under such dynamic channel conditions. This is particularly relevant given the growing demand for high-capacity, long-haul terrestrial FSO systems for next-generation wireless networks. While coherent FSO systems offer considerable potential for such applications, a comprehensive theoretical evaluation of adaptive communication techniques tailored to these scenarios remains largely unaddressed in current literature.

\textcolor{black}{Despite the recent practical advances of adaptive coherent terrestrial FSO communication systems highlighted above, such systems have attracted comparatively limited rigorous theoretical analyses in the literature, as summarized next.} Prior works on IM-DD and coherent HD based terrestrial FSO communication systems have examined limits on the achievable average spectral efficiency (ASE) under lognormal and GG turbulence models (e.g., \cite{gappmair2011further,Hassan2013coherentHD,benkhelifa2013low,ansari2013unified,lee2016effects,Nistazakis2008}, and references therein). More recent studies have extended this analysis to generalized turbulence channels, such as those modeled by the double-generalized gamma and Málaga distributions \cite{doubleGG2015}, \cite{ansari2015}, \cite{osamah2021}, \cite{shishter2024}. Despite these contributions, the crucial aspect of transmitter adaptation has largely been overlooked, with only a few notable exceptions \cite{gappmair2011further}, \cite{Hassan2013coherentHD}, \cite{benkhelifa2013low}. \textcolor{black}{Among these, \cite{Hassan2013coherentHD} investigates a closely related problem, analyzing the ASE of an HD-based adaptive coherent FSO system operating over GG turbulence. However, this analysis neglects the impact of pointing errors, which constitutes a major impairment in practical terrestrial FSO links employing coherent detection.} Further, \cite{ansari2015} derives a unified ASE expression for IM-DD and HD based coherent FSO communication systems operating under the Málaga distribution---subsuming the GG model as a special case---but assumes no transmitter adaptation. Beyond these analytical limitations, several prior studies~\cite{gappmair2011further,ansari2013unified,lee2016effects,doubleGG2015,ansari2015} provide limited clarity regarding the methodology used to model and simulate varying turbulence conditions. In practical terrestrial FSO deployments, key channel parameters are inherently interdependent; for example, variations in link distance simultaneously influence turbulence strength, path loss, and pointing error statistics. A rigorous understanding of these interrelations, grounded in physically consistent channel modeling, is therefore essential for reliable performance evaluation. Neglecting such dependencies may lead to misleading interpretations and flawed insights. Collectively, these observations highlight the absence of a comprehensive baseline spectral efficiency characterization for adaptive coherent terrestrial FSO systems operating under realistic turbulence conditions, pointing errors, and stringent transmit optical power constraints.

Motivated by the above observations, this paper investigates adaptive transmission for terrestrial coherent FSO links with a focus on spectral efficiency optimization under realistic channel impairments. By explicitly exploiting the temporal fluctuations induced by atmospheric turbulence and pointing errors, the proposed analysis aims to characterize the fundamental performance limits of adaptive coherent FSO systems operating under stringent transmit power constraints. The results provide a baseline spectral efficiency characterization that can serve as a reference for the design and evaluation of future adaptive coherent FSO systems subject to practical deployment considerations. The principal contributions of this paper are summarized as follows:
\begin{itemize}
\item We present a novel and exact closed-form expression for the ASE limit of adaptive coherent transmission with synchronous heterodyne (HD) reception over terrestrial FSO links impaired by GG turbulence and pointing jitters. Theorem~\ref{eq:thm2}, together with its specialization in Corollary~\ref{eq:thm3} for the case without pointing errors, constitutes the primary analytical contribution of this work.
%\item To gain insights into the impact of optical turbulence and pointing errors on the adaptive FSO system performance, the exact ASE is further analyzed in the asymptotic regime of high and low optical power levels. Theorem~\ref{eq:theorem4} captures these limiting behaviors of ASE, revealing the associated performance gains and penalties in typical terrestrial FSO links.
\item To provide deeper insights into the impact of turbulence and pointing errors on adaptive FSO system performance, the exact ASE limit is further analyzed in the asymptotic high-SNR and low-SNR regimes. Theorem~\ref{eq:theorem4} rigorously characterizes these limiting behaviors, thereby elucidating both the achievable performance gains and the inherent penalties under practical terrestrial FSO conditions.

\item The validity of the derived ASE limit expressions is rigorously verified through extensive Monte Carlo simulations, which demonstrate excellent agreement across wide and practically relevant ranges of SNR, turbulence strength, and pointing error, as illustrated in Figures \ref{fig:ExWPE}--\ref{fig:PE3}.
\item Among the key findings, two particularly noteworthy and counterintuitive insights are identified:
\noindent i) at high SNRs, the relative ASE degradation diminishes as turbulence strengthens under severe pointing-error conditions (cf. Fig.~\ref{fig:PE3}); and
\noindent ii) at low SNRs, the ASE increases as turbulence strengthens (cf. Fig.~\ref{fig:WPEAsymLowHD}). These effects are rigorously characterized and analytically substantiated.
\end{itemize}
\noindent
\emph{Paper Organization:} The optical turbulence and pointing error models adopted for terrestrial FSO channels are presented in Subsection \ref{subsec:turb_and_jitter}. The relevant noise source and the adaptive coherent FSO system model are described in Subsections \ref{subsec:noises} and~\ref{subsec:detectionschemes}, respectively. Theorems on the ASE limit of the terrestrial coherent FSO channel are derived in Section \ref{sec:CapResults}, and their extensive numerical analysis is presented in Section \ref{sec:numericalR}. Finally, Section \ref{sec:conclude} summarizes the main results and presents concluding remarks and future research directions.

\section{Channel and System Models}\label{sec:system_and_channel}
\subsection{Optical Turbulence with Pointing Errors}\label{subsec:turb_and_jitter}
The overall intensity fluctuation, or fading, $I$, of a transmitted laser beam due to atmospheric turbulence and pointing errors is modeled as
\begin{align}
    I \, = \, I_{l} I_{\mathrm{a}} I_{\mathrm{p}}
\end{align}
where $I_{l}$ represents the path loss component, determined by the exponential Beer-Lambert law for a given link length and weather conditions~\cite{weichel1990laser}, $I_{\mathrm{a}}$ accounts for turbulence-induced fluctuations, and $I_{\mathrm{p}}$ corresponds to pointing-error-induced fluctuations. In this work, we assume $I_{l} = 1$. The statistical models for these independent fading components are described next.

\emph{Gamma-Gamma (GG) Turbulence:} The statistics of the intensity fluctuation $I_{\mathrm{a}}$ across weak-to-strong turbulence regimes are well modeled by the GG distribution, given by
%Under the G–G distribution regime, the signal irradiance $I_{\mathrm{a}}$ is modeled as \cite{andrews1986mathematical}:
\begin{equation}\label{eq:gg_dist}
f_{I_{\mathrm{a}}}( I_{\mathrm{a}}) = \frac{2(ab)^{(a +b)/2}}{\Gamma ( a) \Gamma ( b)}  I_{\mathrm{a}}^{{(a +b)/2 -1}} K_{a -b } \left( 2\sqrt{a b I_{\mathrm{a}}}\right);\ I_{\mathrm{a}}  >0
\end{equation}
where $a$ and $b$ are shape parameters~\cite{GGmodel}, $\Gamma(\cdot)$ is the Gamma function \cite[Eq. (8.310.1)]{gradshteyn2000table}, and $K_v(\cdot)$ is the modified Bessel function of the second kind and order $v$ \cite[Eq. (8.494.1)]{gradshteyn2000table}. For plane-wave optical radiation at the receiver, the GG shaping parameters depend on atmospheric conditions as
%\begin{align}
%   a \,&=\, \left(\exp \left[\frac{0.49\sigma _{{\mathrm{\scriptscriptstyle R}}}^{2}}{\left( 1 + 1.11 \sigma _{{\mathrm{\scriptscriptstyle R}}}^{12/5}\right)^{7/6}}\right] -1\right)^{-1},\label{eq:a}\\
%   b \,&=\,\left(\exp \left[\frac{0.51\sigma _{{\mathrm{\scriptscriptstyle R}}}^{2}}{\left( 1 + 0.69\sigma _{{\mathrm{\scriptscriptstyle R}}}^{12/5}\right)^{5/6}}\right] -1\right)^{-1},\label{eq:b}
%\end{align}
\begin{align}
a \,&=\, \biggl[ \,{\exp \,\biggl( {{0.49\sigma_{\mathrm{\scriptscriptstyle R}}^2}\,} {\Bigl( {1 + 1.11\sigma_{\mathrm{\scriptscriptstyle R}}^{12/5}} \Bigr)}^{-7/6} \biggr) - 1} \,\biggr]^{ - 1},\label{eq:a}\\
b \,&=\,  \biggl[ \,{\exp \, \biggl( {0.51\sigma_{\mathrm{\scriptscriptstyle R}}^2 \,\Bigl( 1+0.69\sigma_{\mathrm{\scriptscriptstyle R}}^{12/5}\Bigr)^{-5/6}}\biggr) - 1}\, \biggr]^{ - 1},\label{eq:b}
\end{align}
where $\sigma _{{\mathrm{\scriptscriptstyle R}}}^{2} = 1.23\,C_{\mathrm{n}}^2 k_{\mathrm{w}}^{7/6}{L^{11/6}}$ is the \emph{Rytov variance} determining the atmospheric turbulence strength, $k_{\mathrm{w}} = 2\pi /{\lambda_{\mathrm{w}}}$ is the optical wave number, ${\lambda_{\mathrm{w}}}$ is the wavelength, $L$ is the propagation distance, and $C_{\mathrm{n}}^2$ is the index-of-refraction structure parameter \cite{GGmodel}. %$C_{\mathrm{n}}^2$ can be practically assumed to be constant for the line-of-sight propagation on horizontal paths~\cite{beland1993}.

\emph{Pointing Error:} Pointing error refers to the misalignment of the laser beam with the optical receiver, caused by i) boresight, the fixed displacement between the beam centroid and the receiver center, and ii) jitter, the random temporal displacement due to factors such as building sway, wind, or vibrations. While careful deployment or fast-tracking transmitters can minimize boresight misalignment, residual jitter remains a concern even with a well-designed ATP subsystem~\cite{andrews2005laser}. To account for this effect, the severity of jitter in this work is later categorized into representative regimes reflecting different levels of ATP capability (see Table~\ref{tab:FSO_System_settings} in Section~\ref{sec:numericalR}).
 
After propagating a distance $L$, the laser beam with waist $w_\mathrm{\scriptscriptstyle L}$ is incident on the photodetector with aperture radius $r_\mathrm{\scriptscriptstyle A}$~\cite{saleh2019fundamentals}. The received optical beam attenuation due to turbulence-induced spread and pointing error, described by radial displacement $r$ between the beam footprint and the detector center, is given by $I_{\mathrm{p}}(r;L) \approx A_\mathrm{\scriptscriptstyle 0} \exp(-2r^2/w_\mathrm{\scriptscriptstyle L_{\mathrm{eq}}}^{2})$ where $w_\mathrm{\scriptscriptstyle L_{\mathrm{eq}}}^{2} =w_\mathrm{\scriptscriptstyle L}^{2}\sqrt{\pi } \,\text{erf}(v) /( 2v\exp( -v^{2}))$ is the (squared) equivalent beam width with $v=\sqrt{\pi} r_\mathrm{\scriptscriptstyle A}/ \sqrt{2} w_\mathrm{\scriptscriptstyle L}$. The fraction of the collected power at $r=0$ is given by 
\begin{align}\label{eq:fractionPowA0}
A_\mathrm{\scriptscriptstyle 0} \,\approx\, \left[\mathrm{erf}(v)\right]^{2},
\end{align}
where $\text{erf}(\cdot)$ is the standard error function \cite[Eq. (8.250.1)]{gradshteyn2000table}, and $0 \leqslant A_\mathrm{\scriptscriptstyle 0} \leqslant 1$~\cite{PEjitter}. For a coherent beam, the beam radius $w_\mathrm{\scriptscriptstyle L}$ is related to the beam waist $w_\mathrm{\scriptscriptstyle 0}$ at the optical transmitter's exit aperture by $\displaystyle w_\mathrm{\scriptscriptstyle L} =w_\mathrm{\scriptscriptstyle 0}( 1+\epsilon ( \lambda_{\mathrm{w}} L/\pi w_\mathrm{\scriptscriptstyle 0}^{2})^2)^{1/2}$ where $\epsilon = ( 1+2w_\mathrm{\scriptscriptstyle 0}^{2} /\rho_{\scriptscriptstyle 0}^{2}(L))$ is the global coherence parameter, $\rho_{\scriptscriptstyle 0}(L) = ( 1.46C_{\mathrm{n}}^{2} k_{\mathrm{w}}^{2} L)^{-3/5}$ is the coherence length in turbulence \cite{ricklin2002}, and the transmit beam waist $w_{\scriptscriptstyle 0}$ is related to the transmitter aperture diameter $D$ as $w_\mathrm{\scriptscriptstyle 0} = {D}/{\sqrt{2}\pi}$ \cite{siegman1986lasers}. Farid and Hranilovic~\cite{PEjitter} derived the PDF of the irradiance component $I_\mathrm{p}$ under the assumptions of i.i.d. zero-mean Gaussian distributions for both horizontal and vertical receiver sways (i.e., zero boresight and identical jitters), expressed as
\begin{equation}\label{eq:pointerrPDF}
\phantom{xxxxxxxxxx}f_{I_{\mathrm{p}}}( I_{\mathrm{p}}) \,=\, \frac{\xi ^{2}}{A_{\scriptscriptstyle 0}^{\xi ^{2}}} I_{\mathrm{p}}^{\xi ^{2} -1}; \ \ \ 0\leqslant I_{\mathrm{p}} \leqslant A_{\scriptscriptstyle 0}\,,
\end{equation}
where the parameter $\xi \triangleq {w_{\mathrm{\scriptscriptstyle L_{eq}}}}/{2\sigma_{\mathrm{e}}}$ is the ratio of the received equivalent beam waist $w_{\mathrm{\scriptscriptstyle L_{eq}}}$ and the standard deviation of the pointing error displacement (jitter). This model is valid when $w_\mathrm{\scriptscriptstyle L}/ r_\mathrm{\scriptscriptstyle A} \gg 1$ and further that the effects of $\xi$ and $A_{\scriptscriptstyle 0}$ parameters are independent \cite{PEjitter}.

\textcolor{black}{A recent refinement of the pointing error model is introduced in~\cite{Miao2023newpointingmodel}, where multiple alternative parameterizations of $\xi$ and $A_{\scriptscriptstyle 0}$ are proposed to enhance modeling accuracy and computational efficiency (see~\cite[Fig.~5]{Miao2023newpointingmodel}). In this work, we hence borrow the \emph{modified intensity uniform model} from~\cite{Miao2023newpointingmodel}, which retains the same mathematical form for the probability distribution as in \eqref{eq:pointerrPDF}, but with redefined parameters given by 
\begin{align}
A_\mathrm{\scriptscriptstyle 0} &\triangleq 1 - \exp(-2 r_\mathrm{\scriptscriptstyle A}^2 / w_\mathrm{\scriptscriptstyle L}^2),\label{eq:redefined_A0}\\
\xi^2 &\triangleq r_\mathrm{\scriptscriptstyle A}^2 /(2 \sigma_{\mathrm{e}}^2 A_\mathrm{\scriptscriptstyle 0}),\label{eq:redefined_xi}
\end{align}
%\begin{flalign}
%&\phantom{xxxxxxxxxxxxx	xxx}A_\mathrm{\scriptscriptstyle 0}  \triangleq   1 \,-\, \exp \biggl[ -\,\dfrac{2 r_\mathrm{\scriptscriptstyle A}^2}{w_\mathrm{\scriptscriptstyle L}^2}\biggr],&\label{eq:redefined_A0}\\
%&\text{and}\phantom{xxxxxxxxxxxxx}\,\xi^2  \triangleq   \dfrac{r_\mathrm{\scriptscriptstyle A}^2} {2 \sigma_{\mathrm{e}}^2 A_\mathrm{\scriptscriptstyle 0}}.&\label{eq:redefined_xi}
%\end{flalign}
Precisely, this choice demonstrates more than an order-of-magnitude reduction in normalized mean square error (NMSE) compared to the Farid–Hranilovic formulation (see~\cite[Table~1]{Miao2023newpointingmodel}), thereby offering a substantially more reliable characterization of pointing jitter effects in terrestrial FSO links.}

\emph{Composite Fading Distribution:} Given the distributions in~\eqref{eq:gg_dist} and~\eqref{eq:pointerrPDF} of the independent constituents $I_{\mathrm{a}}$ and $I_{\mathrm{p}}$, the PDF of the instantaneous composite irradiance $I= I_{\mathrm{a}}I_{\mathrm{p}}$ is developed to~\cite[Eq.~(8)]{gappmair2011further}
\begin{equation}\label{eq:PEPdf}
    f_{I}( I) =\frac{ab\xi ^{2}}{ A_\mathrm{\scriptscriptstyle 0}\, \Gamma ( a) \Gamma ( b)} \, G_{1,3}^{3,0}\left( \frac{abI}{A_\mathrm{\scriptscriptstyle 0}} \,\biggr\rvert\hspace{-0.25em} \begin{array}{c}
\xi ^{2}\\
\xi ^{2} -1, \,a-1,\, b-1\\
\end{array}\hspace{-0.5em}\right).
\end{equation}
\begin{remark}\label{eq:rem1}
For longer terrestrial FSO links, and even for medium-haul links operating under strong turbulence, additional impairments such as boresight misalignment, beam wander, and angle-of-arrival fluctuations become dominant and constitute critical practical challenges. In this work, we focus on turbulence and pointing error effects, an assumption that is well-justified for short-haul links and for medium-haul links under weak turbulence, where higher-order impairments are typically negligible or remain within acceptable limits.
\end{remark}

\subsection{Noise Sources}\label{subsec:noises}
Photodetection in FSO receivers is fundamentally limited by two noise mechanisms: i) shot noise associated with the quantum nature of photoelectric detection~\cite{oliver1965}, and ii) thermal noise originating from the receiver electronics. Shot noise arises due to both the information-bearing optical field and ambient background radiation collected by the receiver aperture, with a variance proportional to the total incident optical power. Importantly, both ambient shot noise and thermal noise are signal-independent, as they remain statistically uncorrelated with the information-carrying optical signal~\cite{agrawal2012fiber}.

In coherent FSO receivers, the received optical field is mixed with a strong optical local oscillator (OLO) signal prior to photodetection. This mixing process introduces an OLO-driven shot-noise component in addition to signal-dependent shot noise, background-induced shot noise, and thermal noise. By appropriately selecting a sufficiently large OLO power, the OLO-induced shot noise can be made to dominate all other noise contributions. Operation in this shot-noise-limited regime is widely regarded as the canonical and practically relevant mode for ideal coherent receivers~\cite{barry90},~\cite{agrawal2012fiber},~\cite{kikuchi2015fundamentals}, and forms the basis for most performance and capacity analyses. By contrast, a direct-detection (DD) receiver produces a photocurrent proportional to the received optical intensity and is subject to signal-dependent shot noise, background-induced shot noise, and thermal noise, with the dominant impairment determined by their relative strengths.

\emph{Remark on the Statistical Characterization of Shot Noise:} Under low received photon counts---as is typical in inter-satellite and ground-to-satellite optical links---the shot noise is accurately modeled using a Poisson distribution~\cite{chaaban2021capacity}. In contrast, for high photon-count regimes, as encountered in most terrestrial FSO applications, a Gaussian approximation is sufficiently accurate. For a rigorous justification of the Gaussian model for high-intensity shot noise, refer~\cite{chaaban2021capacity},~\cite{lee2012digital}.

\subsection{Coherent Heterodyne FSO Receiver Model}\label{subsec:detectionschemes}
%We will now utilize the qualitative description of the FSO receiver noises given in the previous subsection to develop received signal models in IM-DD and synchronous HD schemes.

Let $x_{\mathrm{\scriptscriptstyle HD}} \in \mathbb{C}$  denote the transmitted symbol for the coherent FSO system respectively. We remind the reader that the baseband (information) signal in a coherent modulation scheme represents a complex optical field, whereas the baseband signal in an intensity modulation scheme represents optical intensity. The transmitted optical power is constrained by a fixed long-term constraint denoted by $\mathbb{E}\left[| x_{\mathrm{\scriptscriptstyle HD}}| ^{2}\right] \leqslant P_{\mathrm{avg}}$. \textcolor{black}{As discussed in Section~\ref{sec:intro}, owing to millisecond-scale coherence times and coherence bandwidths on the order of hundreds of gigahertz, a frequency-flat block-fading model is appropriate for terrestrial FSO channels subject to time-varying irradiance fluctuations: each fading state is assumed to span an identical duration corresponding to a fixed coherence time, with i.i.d. fading realizations across blocks. The block-wise channel state is assumed to be quasi-ideally known at both the transmitter and the receiver. In practice, the channel state is estimated at the receiver once at the beginning of each coherence block and conveyed to the transmitter via a feedback link. While channel estimation, processing, and CSI feedback inevitably incur non-zero latency, the quasi-ideal CSI assumption adopted here requires that the overall feedback delay be sufficiently small---on the order of sub-millisecond latency---so as to remain negligible relative to the channel coherence time, thereby enabling timely block-wise signal adaptation at the transmitter.} Adhering to this understanding, we develop the received signal model for the coherent FSO communication system employing HD detection. Assuming ideal synchronous HD detection, the received complex baseband symbol is described by
\begin{equation}
\label{eq:HD}
y_{\mathrm{\scriptscriptstyle HD}} \, = \, h\,x_{\mathrm{\scriptscriptstyle HD}}\, +\,\,w_{\mathrm{\scriptscriptstyle OLO}}\,,
\end{equation}
where $y_{\mathrm{\scriptscriptstyle HD}} \ \in \ \mathbb{C}$ represents its output, $ h \in \mathbb{C}$ is the fluctuation in the received optical field and $w_{\mathrm{\scriptscriptstyle OLO}} \sim  \mathcal{CN}( 0,\sigma ^{2}_{\mathrm{\scriptscriptstyle OLO}})$ is the OLO-induced shot noise. \textcolor{black}{As noted earlier, the OLO shot noise is the only dominant noise of consequence, and the total AWGN from all sources other than the OLO is negligible~\cite[Ch.~3]{karp1988optical}. For the baseband model in~\eqref{eq:HD}, the received electrical SNR conditioned on $h$ is\vspace{-0.5em}
\begin{align}\label{eq:SNR_HD}
\frac{I P(I)}{\sigma_{\mathrm{\scriptscriptstyle OLO}}^{2}}\,\cdot
\end{align}
Here, $ I \triangleq |h|^{2}$ denotes the channel (intensity) gain, and $P(I) \triangleq \mathbb{E}\!\left[ |x_{\mathrm{\scriptscriptstyle HD}}|^{2} \,\big|\, I \right]$ represents the transmit optical power allocated according to the channel state $I$.} 

\textcolor{black}{Since the average transmit power budget and the OLO shot-noise variance are fixed system parameters, it is convenient to express power in normalized form by scaling $P_{\mathrm{avg}}$ with $\sigma_{\mathrm{\scriptscriptstyle OLO}}^{2}$, a ratio that naturally arises in the subsequent derivation.}
\textcolor{black}{\begin{definition}
The normalized transmit power is defined as
\begin{align}\label{eq:scaledPower}
\rho \triangleq \frac{P_{\mathrm{avg}}}{\sigma_{\mathrm{\scriptscriptstyle OLO}}^{2}}\, \cdot
\end{align}
Under ideal coherent FSO conditions---including unit channel gain (i.e., no turbulence or pointing errors and full beam capture) and ideal optical-to-electrical conversion---the entire transmitted optical power is recovered at the receiver, and $\rho$ coincides with the received electrical SNR. 
\end{definition}}

\vspace{-1.7em}
\textcolor{black}{
\begin{remark}
The adopted coherent FSO framework for \eqref{eq:HD} assumes that the OLO-driven receiver shot noise is the dominant impairment. Although transmitter-induced noise sources (e.g., laser RIN, modulation noise, and optical amplifier ASE noise) originate at the transmitter, their effective contribution must be assessed at the receiver after propagation, optical mixing and detection. The present analysis assumes that these contributions are small compared to the OLO shot noise at the receiver, and are therefore not explicitly modeled. Accordingly, the parameter $\rho$ represents a nominal transmit-power normalization referenced to the OLO shot-noise variance at the receiver, and should not be interpreted as a hardware-constrained transmit SNR.
\end{remark}}

\noindent
\textcolor{black}{Any performance variations arise exclusively from the channel statistics and their exploitation via channel-dependent power adaptation observing the budget $\mathbb{E}[P(I)] = P_{\mathrm{avg}}$. Accordingly, $\rho$ is adopted throughout this paper as the channel-independent system resource parameter for ASE evaluation and comparison across different turbulence and pointing-error regimes.}

\section{Achievable Spectral Efficiency With Optimal Adaptive Coherent FSO Transmission}\label{sec:CapResults}
In IM-DD FSO systems, information is mapped onto the intensity of the transmitted optical field, which is inherently non-negative. In constrast, the baseband symbol $ x_{\mathrm{\scriptscriptstyle HD}}$ in~\eqref{eq:HD} utilizes two degrees of freedom. 
Since $w_{\mathrm{\scriptscriptstyle OLO}}$ is AWGN, the channel conditioned on the fading state admits the classical Shannon capacity expression. Accordingly, for the instantaneous received SNR in~\eqref{eq:SNR_HD}, the achievable spectral efficiency (in nats/s/Hz) of the coherent FSO channel in~\eqref{eq:HD} is
\begin{align}\label{eq:CapInt}
\mathrm{S}_{\scriptscriptstyle \mathrm{HD}} (\lambda) \, &= \,\,\ln \left( 1 + \dfrac{ \lambda  P (\lambda)}{\sigma_{\mathrm{\scriptscriptstyle OLO}}^{2}}\right)\,\centredComma
\end{align}
where $\lambda \triangleq I$ denotes the block fading state.
\noindent
The maximum long-term average (over blocks) or simply average spectral efficiency (ASE) is defined as
\begin{align}\label{eq:capacity}
\widebar{\mathrm{S}}_{\scriptscriptstyle \mathrm{HD}} \, \triangleq \,\,\,\max_{P} \,\,\,&\,  \mathbb{E}\left[\ln\left( 1 + \frac{\lambda P (\lambda)}{\sigma_{\mathrm{\scriptscriptstyle OLO}}^{2}}\right) \right]\,\centredComma
\end{align}
where the maximization is performed over all feasible beam adaptations choices of $P$ that satisfy $\mathbb{E}[P (\lambda)] \leqslant P_{\mathrm{avg}}$. 

\noindent
The optimal beam power adaptation $P^{\,*}$ maximizing~\eqref{eq:capacity} is
\begin{align}\label{eq:waterfilling}
\dfrac{P^{\,*} (\lambda)}{\sigma_{\mathrm{\scriptscriptstyle OLO}}^{2}} \,=\,\left(\dfrac{1}{\mu} -\dfrac{1}{\lambda}\right)^{+}\,\centredComma
\end{align}
with $\mu$ chosen so that the (normalized) optical power constraint
\begin{align}\label{eq:powerint}
\mathbb{E} \biggl[\left(\dfrac{1}{\mu} -\dfrac{1}{\lambda}\right)^{+}\biggr] \,=\, \rho
\end{align}
is satisfied. Additionally, the parameter $\mu$ serves as a channel cutoff, which simplifies the ASE problem in \eqref{eq:capacity} to
\begin{align}\label{eq:CapInt234}
\widebar{\mathrm{S}}_{\scriptscriptstyle \mathrm{HD}} \,&=\,\,\int_{\mu}^{\infty}   \ln\left(\dfrac{\lambda}{\mu}\right) f_{\lambda}(\lambda) d\lambda.
\end{align}
The integral in~\eqref{eq:CapInt234} can be divided into three parts as follows:
\begin{align}\label{eq:integral_bifurc}
&\widebar{\mathrm{S}}_{\scriptscriptstyle \mathrm{HD}} \,=\, \, \Biggl[\,\,\underbrace{\int _{0}^{\infty }  \!\! \ln( \lambda )f_{\lambda }( \lambda ) d\lambda \ }_{= \, I_{1}} \,\,\,-\,\phantom{xxxxxxxxxxxxxxxxxxxxxxx} \notag\\
&\phantom{xxxxx}\underbrace{\int _{0}^{\mu}  \!\! \ln( \lambda ) f_{\lambda }( \lambda ) d\lambda \ }_{=\,I_{2}} \,-\,\underbrace{\int _{\mu}^{\infty }  \!\! \ln( \mu) f_{\lambda }( \lambda ) d\lambda }_{=\,I_{3}}\,\, \Biggr].
\end{align}
\noindent
In evaluating the $I_1, \, I_2$ and $I_3$ integrals, we will frequently apply an integral identity involving the Meijer-G function \cite[Eq.~(9.301)]{gradshteyn2000table}, which is reproduced below (see \cite{wolframMeijerGfunction} for additional details).
\begin{flalign}\label{eq:taildisb22}
&\int \!\!x^{j - 1} G_{p,q}^{m,n}\left(x\,
\middle\vert\hspace{-0.15em}
\begin{array}{c}
a_1,\ldots,,a_p\\
b_1,\ldots,,b_q
\end{array}\hspace{-0.5em}
\right) dx \,=\phantom{xxxxxxxxxxxxxxxxxxxxxxxxxxxx}&\notag\\
&G^{m, \,n+1}_{\!p+1,q+1}\hspace{-0.25em}\left(x\middle\vert\hspace{-0.4em}
\begin{array}{c}
1, j\! + a_1,\ldots,j\! +a_n,j \!+a_{n+1},\ldots,j\! + a_p\\
j \!+ b_1,\ldots,j\! + \!b_m,\!0,j\!+b_{m+1},\ldots,j\!+b_q
\end{array}\hspace{-0.5em}
\right)\!\cdot &
\end{flalign}

\subsection{Exact ASE With and Without Pointing Errors}\label{subsec:exactC}
\begin{theorem}\label{eq:thm2}
For the adaptive coherent terrestrial FSO channel in~\eqref{eq:HD}, with intensity gain $I = |h|^2$ modeled by a composite fading distribution accounting for GG turbulence and pointing errors as characterized in~\eqref{eq:PEPdf}, and assuming perfect channel state information (CSI) at both the transmitter and the receiver, the exact ASE in \eqref{eq:capacity} simplifies to
\begin{align}\label{eq:ExactPE}
\widebar{\mathrm{S}}_{\scriptscriptstyle \mathrm{HD}} \,&=\,\, \biggl[\,\ln\left(\frac{A_\mathrm{\scriptscriptstyle 0}}{ab}\right) + \,\psi (a) + \,\psi (b) -\,\ln(\mu) - \left(\frac{1}{\xi ^{2}}\right)\notag \\
\,&\phantom{x}\,\,\,\,+\, \frac{\xi ^{2}}{\Gamma ( a) \Gamma ( b)} \, G_{3,5}^{3,2}\left(\frac{ab}{A_\mathrm{\scriptscriptstyle 0}}\mu\,\biggr\rvert\hspace{-0.25em} \begin{array}{c}
1,\,1,\,\xi ^{2}+1\\[0.1em]
\xi ^{2}, \,a,\, b,\,0,\,0\\
\end{array}\hspace{-0.5em}\right)\,\biggr]
\end{align}
where $\psi(\cdot)$ is the Euler's digamma function, and $\mu$ is determined from $\int_{\mu }^{\infty} (1/\mu - 1/\lambda) f_{{\lambda}} (\lambda) d \lambda \,=\, \rho$.
\end{theorem}
\begin{IEEEproof}
\noindent
The exact ASE expression (from~\eqref{eq:integral_bifurc}) is 
\begin{align}\label{eq:cap_intparts}\vspace{-1em}
\widebar{\mathrm{S}}_{\scriptscriptstyle \mathrm{HD}} \,&=\, I_1 \,-\, I_2 \,-\, I_3.
\end{align}
\underline{\emph{Evaluation of Integral $I_1$}}: The integral $I_1$ can be viewed as
\begin{equation}\label{eq:I1_1}
    I_{1} \,=\, \mathbb{E}\left[\ln \lambda \right].
\end{equation}
We reformulate \eqref{eq:I1_1} by taking expectation on the identity $\frac{d}{dt} \lambda^{t} =\lambda ^{t} \ln (\lambda)$ and thereupon evaluating at $t = 0$ such that
\begin{flalign}\label{eq:k_th_moment}
&\phantom{xxxxxxxxxxxx}\mathbb{E}\left[ \lambda ^{t} \ln( \lambda )\right]\bigl{|}_{\,t = 0} \,\, = \, \frac{d}{dt} \left( \mathbb{E}\left[\lambda^{t}\right]\right)\biggl{|}_{\,t=0}\,\,, &\phantom{xxxxxxxxxxxxxxx}\notag\\
&\text{and hence}\phantom{xxxxx}  I_{1} \, =\, \frac{d}{dt}\left(\int _{0}^{\infty } \!\!\lambda^{t} f_{\lambda }( \lambda )d\lambda \right)\biggl{|}_{\,t = 0} \,\,.&
\end{flalign} 
The $t$-th moment in the RHS of~\eqref{eq:k_th_moment} is computed as
\begin{flalign}\label{eq:kmoment1}
&\int _{0}^{\infty}  \!\!\lambda^t f_{\lambda }( \lambda )d\lambda \,\,=\notag &\\
& \frac{ab\xi ^{2}}{ A_\mathrm{\scriptscriptstyle 0} \Gamma ( a) \Gamma ( b)} \int _{0}^{\infty}  \!\! \lambda^t   G_{1,3}^{3,0} \left( \frac{ab \lambda}{A_\mathrm{\scriptscriptstyle 0}}  \biggr\rvert\hspace{-0.25em} \begin{array}{c}
\xi ^{2}\\
\xi ^{2}  \!-1,  \!a-1, \!b-1
\end{array}\hspace{-0.45em}\right)  d\lambda.&
\end{flalign}
Applying~\eqref{eq:taildisb22} in~\eqref{eq:kmoment1}, and making the substitutions:
\begin{flalign}
&\phantom{x}\,\,\lim_{x \to 0}\,\,G_{2,4}^{3,1}\!\left(x\biggr\rvert\hspace{-0.25em} \begin{array}{c}
 1,\xi^{2}\!+t+1\\
\!\xi^{2}\!+t,\!a\!+t,\!b\!+t,\!0\\
\end{array}\hspace{-0.5em}\right) = 0,\,\,\,\text{and}&\\
&\phantom{x}\,\,\lim_{x \to \infty}\,G_{2,4}^{3,1}\!\left(x\biggr\rvert\hspace{-0.25em} \begin{array}{c}
 1,\xi^{2}\!+t+1\\
\!\xi ^{2}\!+t,\!a\!+t,\!b\!+t,\!0\\
\end{array}\hspace{-0.5em}\right) = \frac{\Gamma \!\left( a \!+ t\right) \!\Gamma\! \left( b\!+ t \right)}{\xi^{2}\! +t}\,\centredComma&
\end{flalign}
we get
\begin{equation}
 I_{1} \,=\,\frac{\xi ^{2}}{\Gamma\!(a)\Gamma\!(b)} 
 \, \frac{d}{dt}\left(\left(\frac{A_\mathrm{\scriptscriptstyle 0}}{ab} \right)^{\!t}\frac{\Gamma \left( a+ t\right) \Gamma \left( b + t\right)}{\xi ^{2} + t}\right) \Biggl{|}_{\,t = 0}.   
\end{equation}
Upon differentiation w.r.t. $t$ and then evaluating the attained expression for $t =
0$, we obtain the final $I_{1}$ expression as
\begin{equation}
 I_{1} \ =\  \ln\left(\frac{A_\mathrm{\scriptscriptstyle 0}}{ab}\right) + \psi (a) +\psi (b) -\left(\frac{1}{\xi ^{2} }\right) \,
\cdot
\end{equation}
In the above, we have used $d(\ln\Gamma(x))/dx \,=\, \psi(x)$.

\noindent
\underline{\emph{Evaluation of Integral $I_2$}}: The integral part $I_{2}$ is now evaluated as
\begin{align}
&I_2 = \int _{0}^{\mu}  \!\! \ln( \lambda ) f_{\lambda }( \lambda )d\lambda \phantom{xxxxxxxxxxxxxxxssssssssssssxx}\,\notag \\
&= \frac{ab\xi ^{2}}{ A_\mathrm{\scriptscriptstyle 0} \Gamma ( a) \Gamma ( b)} \int _{0}^{\mu} \ln\left(\lambda\right) G_{1,3}^{3,0}\left(\hspace{-0.1em} \frac{ab \lambda}{A_\mathrm{\scriptscriptstyle 0}}\biggr\rvert\hspace{-0.45em} \begin{array}{c}
\xi^{2}\\[0.1em]
\xi^{2}-1,a-1,b-1\\
\end{array}\hspace{-0.55em}\right)\hspace{-0.15em} d\lambda.\notag
\end{align}
Applying integration by parts with $\ln(\cdot)$ as the first function and the Meijer-G function as the second yields
%\fontsize{9.7}{12}\selectfont
\begin{align}
I_2 \, &= \, \biggl[G_{2,4}^{3,1}\left(\dfrac{ab\lambda}{A_\mathrm{\scriptscriptstyle 0}}\,\biggr\rvert\hspace{-0.25em}  \begin{array}{c}
  1, \xi ^{2} +1 \\[0.1em]
\xi ^{2}, a,  b, 0
\end{array}\hspace{-0.35em}\right) \times \ln(\lambda)\notag \\
&\,\,\,\,\,\,\,\,\,- G_{3,5}^{3,2}\left(\dfrac{ab\lambda}{A_\mathrm{\scriptscriptstyle 0}} \, \biggr\rvert\hspace{-0.25em}  \begin{array}{c}
 1, 1, \xi ^{2} +1 \\[0.1em]
\xi ^{2}, a, b, 0, 0
\end{array}\hspace{-0.35em}\right)\biggr]\, \biggl{|}_{\,\lambda = 0}^{\,\lambda = \mu} \,\,\times\,\,\,\dfrac{\xi ^{2}}{\Gamma ( a) \Gamma ( b)} \, \cdot
\end{align}  
%\normalfont
With the Meijer-G function limits:
%\fontsize{9.7}{12}\selectfont
\begin{flalign}
&\phantom{xxxxxx}\lim_{x \to 0} \,\,G_{2,4}^{3,1}\left(x\,\biggr\rvert\hspace{-0.25em} \begin{array}{c}
  1, \xi ^{2} +1 \\[0.1em]
\xi ^{2}, a,  b, 0
\end{array}\hspace{-0.35em}\right) \,\times\,\ln(x)\,=\, 0,\,\,\,\text{and} &\\
&\phantom{xxxxxx}\lim_{x \to 0} \,\,G_{3,5}^{3,2}\left(x \,\biggr\rvert\hspace{-0.25em}\begin{array}{c}
 1, 1, \xi ^{2} +1 \\[0.1em]
\xi ^{2}, a, b, 0, 0
\end{array}\hspace{-0.35em}\right)\,=\, 0,&
\end{flalign}
%\normalfont
the final $I_2$ expression is given by
%\fontsize{9.7}{12}\selectfont
\begin{align}
I_{2} \,&=\, \left[ G_{2,4}^{3,1}\left(\dfrac{ab}{A_\mathrm{\scriptscriptstyle 0}}\mu\,\biggr\rvert\hspace{-0.25em}  \begin{array}{c}
  1, \xi ^{2} +1 \\[0.1em]
\xi ^{2}, a,  b, 0
\end{array}\hspace{-0.35em}\right)\right. \times \,\ln(\mu)\notag \\[0.15em]
\,&\,\,\,\,\,\,\,\,\,- \left. G_{3,5}^{3,2}\left(\dfrac{ab}{A_\mathrm{\scriptscriptstyle 0}}\mu \,\biggr\rvert\hspace{-0.25em} \begin{array}{c}
 1, 1, \xi ^{2} +1 \\[0.1em]
\xi ^{2}, a, b, 0, 0
\end{array}\hspace{-0.35em}\right)\right] \,\,\times\,\,\, \dfrac{\xi ^{2}}{\Gamma ( a) \Gamma ( b)} \, \cdot\,\,\,\,
\end{align}
\underline{\emph{Evaluation of Integral $I_{3}$}}:
The $I_3$ component is given as
\begin{align}
I_{3} \,&=\, \ln( \mu) \int _{\mu}^{\infty}   \!\!f_{\lambda }( \lambda ) d\lambda,
\end{align}
where the integral is computed as
\begin{align}
&\int_{\mu}^{\infty} \!\!f_{\lambda }( \lambda ) d\lambda\phantom{xxxxxxxxxxxxxxxxxxxxxxxxxxxxxxxxx}\notag\\
&=\, 1 - \frac{ab\xi ^{2}}{ A_\mathrm{\scriptscriptstyle 0} \Gamma ( a) \Gamma ( b)} \int ^{\mu}_{0}   \!\! G_{1,3}^{3,0}\left( \frac{ab\lambda}{A_\mathrm{\scriptscriptstyle 0}} \,\biggr\rvert\hspace{-0.25em} \begin{array}{c}
\xi ^{2}\\[0.1em]
\xi ^{2} -1,a-1, b-1\\
\end{array}\hspace{-0.5em}\right)d\lambda\notag \\
\,&= \, 1 - \frac{\xi ^{2}}{ \Gamma ( a) \Gamma ( b)} \, G_{2,4}^{3,1}\left( \frac{ab}{A_\mathrm{\scriptscriptstyle 0}}\mu \,\biggr\rvert\hspace{-0.25em}\begin{array}{c}
 1,\,\xi^{2}+1\\[0.1em]
\xi ^{2},\,a,\,b,\,0\\
\end{array}\hspace{-0.35em}\right) \, \cdot
\end{align}
The last equality is obtained by applying the identity in~\eqref{eq:taildisb22} along with the substitution that
\begin{align*}
\lim_{x \to 0}\,\,G_{2,4}^{3,1}\left(x\,\biggr\rvert\hspace{-0.25em} \begin{array}{c}
 1,\,\xi^{2}+1\\[0.15em]
\xi ^{2},\,a,\,b,\,0\\
\end{array}\hspace{-0.35em}\right) \,=\, 0.    
\end{align*}
Substituting these $I_1$,\,$I_2$ and $I_3$ expressions back into~\eqref{eq:cap_intparts} and consequent simplification gives the final closed-form expression of the ASE in Theorem~\ref{eq:thm2}.
\end{IEEEproof}

The exact ASE performance with optimal beam power adaptation over the GG turbulence channel, excluding pointing errors, serves as a useful baseline for comparison.
\begin{cor}\label{eq:thm3}
For the adaptive coherent terrestrial FSO channel in~\eqref{eq:HD}, with intensity gain $I = |h|^2$ modeled by a GG distribution without pointing errors as characterized in~\eqref{eq:gg_dist}, and assuming perfect channel state information (CSI) at both the transmitter and receivers, the exact ASE is given by
\begin{align}
\label{eq:ExactwithoutPE}
\phantom{xxx}\widebar{\mathrm{S}}_{\scriptscriptstyle \mathrm{HD}} \,=\,\, &\biggl[\, \ln\left(\frac{ 1}{ab}\right) \,+ \,\, \psi (a) \,+\,\, \psi (b) \,-\,\,\ln\left(\mu\right)\notag \\
&+\, \frac{1}{\Gamma (a) \Gamma ( b)}  \,G_{2,4}^{2,2}\left({ab}\mu\,\biggr\rvert\hspace{-0.25em} \begin{array}{c}
1,1\\[0.1em]
a, b,0,0\\
\end{array}\hspace{-0.35em}\right)\,\biggr]&
\end{align}
where $\mu$ is solved from $\int_{\mu }^{\infty} (1/\mu - 1/\lambda) f_{{\lambda}} (\lambda) d \lambda \,=\, \rho$.
\end{cor}
\begin{IEEEproof} In the limiting case where $A_\mathrm{\scriptscriptstyle 0} \!\to\! 1$ and $\xi^2  \!\to\! \infty$, the composite fading distribution in~\eqref{eq:PEPdf} converges to the GG distribution in \eqref{eq:gg_dist}. Substituting these into~\eqref{eq:ExactPE} and applying the identity from~\cite[Eq.~(9.31.1)]{gradshteyn2000table} completes the proof.
%Alternatively, the capacity for the GG turbulence channel `without pointing error' case can be solved using a similar approach discussed in  Section~\ref{sec:ExactCapacityProof} but with~\eqref{eq:gg_dist} as the chosen distribution.
\end{IEEEproof}
To our knowledge, Theorem~\ref{eq:thm2} and Corollary~\ref{eq:thm3} provide novel exact ASE expressions for adaptive coherent terrestrial FSO links under GG turbulence, with and without pointing errors.
%Limitations of the $\ln(\mu)$ term and the Meijer-G based term can be overcome by noting that both of these terms converge in the extreme SNR regimes.

\subsection{Immediate Insights}
The proposed ASE limit is interesting in obtaining quantitative as well as qualitative insights as follows.
\begin{itemize}
\item At high SNRs, contribution from the Meijer-G based term vanishes while $\mu \to 1 / \rho$. Keeping this in mind, and now comparing the ASE of channels with and without pointing error, the ASE loss due to pointing jitters can be easily quantified in terms of $A_\mathrm{\scriptscriptstyle 0}$ and $\xi^2$ parameters.
\item Likewise, the ASE loss at high SNRs due to both turbulence and pointing error when compared with no channel impairments (pure AWGN) can also be quantified.
\item In a given terrestrial FSO link configuration, variations in turbulence statistics can influence the associated pointing error characteristics. The derived closed-form ASE expression enables accurate assessment of the coupled impact of these interdependent channel parameters.
\end{itemize}
A more in-depth and broader exposition of the impact of optical channel impairments and adaptive transmission benefits will be presented in Section~\ref{sec:numericalR}, allowing for a wide range of variations in turbulence, pointing error, and SNR.
%These results provide some insight on the impact of the fading and pointing error parameters on the FSO systems performances except for the $\ln(\mu)$ term and the Meijer-G based term; both of these terms converge in the extreme SNR regimes as explained next. %Even though the capacity values can be directly computed in this case, the asymptotic analysis provides much insight as to how exactly the fading statistics and pointing error impairments affect the performance.
% \begin{figure*}[!htbp]
% \begin{equation}
% C_{Exact} =-\dfrac{1}{\xi ^{2}} +\ln\left(\frac{A_\mathrm{\scriptscriptstyle 0} \Omega _{p}}{ab}\right) +\psi ( 0,a) +\psi ( 0,b) \\
% +\left(\frac{\xi ^{2}}{\Gamma ( a) \Gamma ( b)} G_{35}^{32}\left(\frac{ab}{A_\mathrm{\scriptscriptstyle 0}}\left(\frac{\lambda _{c}}{\Omega _{p}}\right) |\ \begin{matrix}
%  & 1 & 1 & \xi ^{2} +1 & \\
% \xi ^{2} & a & b & 0 & 0
% \end{matrix}\right)\right) - \ln( \lambda _{c})
% \end{equation}
% \end{figure*}
% \begin{theorem}
% \label{thm1}
% For a Gamma-Gamma faded FSO communication channel under an
% average-power constraint \eqref{eq:CapInt} with perfect CSI at both the transmitter and the
% receiver side, the exact capacity without pointing error is given by
% \end{theorem}
% \begin{multline}
%  \label{eq:ExactNPE}
% C_{Exact} =\ln\left(\frac{\Omega _{p}}{ab}\right) +\psi ( 0,a) +\psi ( 0,a) \\
% +\frac{1}{\Gamma ( a) \Gamma ( b)}\left( G_{24}^{22}\left( ab\left(\frac{\lambda _{c}}{\Omega _{p}}\right)\Bigl| \ \begin{matrix}
%  & 1 & 1 & \\
% a & b & 0 & 0
% \end{matrix}\right)\right) -\ln( \lambda _{c})

\subsection{Low-SNR and High-SNR Asymptotic Expressions}\label{subsec:AsympC} 
The asymptotic expansions of the exact ASE expression should further reveal how key fading channel parameters from the GG turbulence and the pointing error distributions influence the performance of adaptive FSO schemes. To this end, the $\mu$–$\rho$ relationship is first derived under average power constraint in both low-SNR and high-SNR regimes.
\begin{lemma}\label{eq:lemma_threshold}
The threshold $\mu$ corresponding to the optimal beam power adaptation in~\eqref{eq:waterfilling} satisfies
\begin{flalign}
&\mathrm{(i)}\,\,\,\text{At high SNR:} \,\, \mu \, \approx \, \dfrac{1}{\rho},&\\
&\mathrm{(ii)}\,\,\text{At low SNR:} \,\,\, \mu \, \approx \,  \dfrac{A_\mathrm{\scriptscriptstyle 0}}{4ab} \ln^2 \left(\dfrac{1}{\rho} \right).&
\end{flalign}
\end{lemma}
\begin{IEEEproof}
\textit{Part} $(i)$: Substituting~\eqref{eq:PEPdf} into the average optical transmit power constraint~\eqref{eq:powerint}, we have
%\fontsize{9.7}{12}\selectfont
\begin{align*}
\dfrac{\rho \mu }{\mathcal{B}} =
\int_{\mu}^{\infty}  \biggl(1 - \dfrac{\mu}{\lambda} \biggr)\, G_{1,3}^{3,0}\left( \frac{ab}{A_\mathrm{\scriptscriptstyle 0}}\lambda\, \biggr\rvert\hspace{-0.25em} \begin{array}{c}
\xi^{2}\\[0.1em]
\xi^{2}-1,a-1,b-1\\
\end{array}\hspace{-0.45em}\right)d\lambda\,, \notag
\end{align*}
%\normalfont
where $\mathcal{B} := {ab \xi^{2}/ (A_\mathrm{\scriptscriptstyle 0} \Gamma ( a) \Gamma ( b)})$. Applying the identity in~\eqref{eq:taildisb22} to the above power constraint equation gives
%\fontsize{9.7}{12}\selectfont
\begin{align}\label{eq:powerintExpanded}
\dfrac{\rho \mu }{\mathcal{B}}  \,=\, \dfrac{1}{\mathcal{B}} \,-\,   & \dfrac{A_\mathrm{\scriptscriptstyle 0}}{ab} G_{2,4}^{3,1}\left(\frac{ab}{A_\mathrm{\scriptscriptstyle 0}}\mu \,\biggr\rvert\hspace{-0.25em}\begin{array}{c}
 1,\,\xi^{2}+1\\[0.1em]
\xi ^{2},\,a,\,b,\,0\\
\end{array}\hspace{-0.35em}\right)\notag\\
 +\, \,\mu \, &G_{2,4}^{3,1}\left(\dfrac{ab}{A_\mathrm{\scriptscriptstyle 0}}\mu\,\biggr\rvert\hspace{-0.25em}  \begin{array}{c}
  1, \xi ^{2} \\[0.1em]
\xi ^{2} -1, a-1,  b-1, 0
\end{array}\hspace{-0.45em}\right).
\end{align}
%\normalfont
From the power constraint in~\eqref{eq:powerint}, it follows that $\mu$ exhibits a strictly monotonic inverse relation with $\rho$, decreasing as $\rho$ increases. Notably, at high SNRs, the contributions of the Meijer-G function terms in~\eqref{eq:powerintExpanded} become negligible due to their vanishing behavior for small arguments. Hence
$$
\displaystyle \lim_{\mu \to\, 0} \,\,\,  \rho \mu \,\,= \,\,1.
$$
This completes the proof of {(i)}.\\ 
$\textit{Part (ii):}$ We next derive an explicit $\mu$–$\rho$ relationship in the low-SNR regime. Since the cutoff $\mu$ increases as $\rho$ decreases, power allocation is restricted to large channel gains. To this end, the channel distribution $f_{\lambda}(\lambda)$ in~\eqref{eq:PEPdf} is approximated using a low-order series expansion of the Meijer $G$-function for large arguments, as detailed below:
\begin{align}\label{eq:G_dist_approx}
G_{1,3}^{3,0}\left(z \,\biggr\rvert\hspace{-0.25em} \begin{array}{c}
\xi ^{2}\\
\xi ^{2} -1,a-1,b-1\\
\end{array}\hspace{-0.35em}\right) \,\approx\,  \dfrac{ \sqrt{\pi}\,  z^{(-7+2a+2b)/4}}{e^{2\sqrt{z}}}\, \cdot
\end{align}
Only the dominant term of the series expansion is shown in~\eqref{eq:G_dist_approx}, and substituting it into the LHS of~\eqref{eq:powerint} yields
\begin{align}\label{eq:cutoff121}
\rho  \,&\approx \,   \mathcal{K} \left(\dfrac{B_1 (ab\mu / A_\mathrm{\scriptscriptstyle 0})}{ab\mu / A_\mathrm{\scriptscriptstyle 0}}  - B_2 (ab\mu / A_\mathrm{\scriptscriptstyle 0}) \right)\,\centredComma
%\dfrac{\sqrt{\pi}}{\Gamma(m_T)\Gamma(m_R)} \int_{\mu_0}^{\infty}\left( \dfrac{1}{\mu_0} -\dfrac{1}{\lambda}\right)  \lambda^{\frac{t %m_T + r m_R}{2} - \frac{5}{4}} \, e^{-2\sqrt{\lambda}}\,\, \mathrm{d} \lambda\\
%&= \dfrac{\sqrt{\pi}}{\Gamma(t m_T)\Gamma(r m_R)} \left[ \dfrac{1}{\mu_0} I_1 (\mu_0) - I_2 (\mu_0) \right]
\end{align}
where $\mathcal{K} := (\sqrt{\pi} \,  a b \,\xi^{2})/(A_\mathrm{\scriptscriptstyle 0}\, \Gamma(a)\,\Gamma(b))$ is a positive constant independent of the threshold value, and
\begin{align}\label{eq:I_definitions}
B_1 (x) \,:=\,\,2^{-n}\,\,\,{\Gamma\bigl(n + 1, 2\sqrt{x}\bigr)},\\
B_2 (x) \,:=\,\,2^{2-n}\,{\Gamma\bigl(n - 1, 2\sqrt{x}\bigr)},
\end{align}
\noindent
where, in turn, $n := a + b - \tfrac{5}{2}$, and $\Gamma(\cdot,\cdot)$ denotes the upper incomplete Gamma function, whose large-argument series expansion is given below~\cite{gradshteyn2000table}:
%\begin{align}\label{eq:gamma_approx}
%\Gamma(s,t) \, \approx \,t^s e^{-t} \bigl[{t}^{-1} + {(s-1)}{t^{-2}} + o({t}^{-3})\bigr].
%\end{align}
\begin{align}\label{eq:gamma_approx}
\Gamma(s,t) \, \approx \, t^s e^{-t}  [\,{t}^{-1} \,+\, {(s-1)}{t^{-2}} \,+\, O({t}^{-3})].
\end{align}
By retaining the two leading terms in~\eqref{eq:gamma_approx},~\eqref{eq:cutoff121} simplifies to
\begin{equation}\label{eq:SNRb}
\rho \,\approx\, \mathcal{K}  \left({ab \mu}/{A_\mathrm{\scriptscriptstyle 0}}\right)^{\left({a+b} - \frac{11}{2}\right)\frac{1}{2}} e^{-2\sqrt{{ab \mu}/{A_\mathrm{\scriptscriptstyle 0}}}}.
\end{equation}
Taking the logarithm of both sides of~\eqref{eq:SNRb} and retaining the dominant term on the right-hand side, we obtain
\begin{align}\label{eq:n_below1223}
\ln \rho \,\approx\, -2\sqrt{{ab \mu}/{A_\mathrm{\scriptscriptstyle 0}}}.
\end{align}
Solving~\eqref{eq:n_below1223} for $\mu$, we finally arrive at
\begin{align}\label{eq:n_below1224}
\mu \,\approx\, \dfrac{A_\mathrm{\scriptscriptstyle 0}}{4ab}\ln^{2}\left(\frac{1}{\rho}\right)\,\cdot \vspace{-1em}
\end{align}
This completes the proof of (ii).
\end{IEEEproof}
\begin{theorem}\label{eq:theorem4}
For the adaptive coherent terrestrial FSO channel in~\eqref{eq:HD}, with intensity gain $I = |h|^2$ distributed according to \eqref{eq:PEPdf} and perfect channel state information (CSI) at both the transmitter and receiver, the asymptotic ASEs at low and high SNR are given by
\begin{flalign}
&\mathrm{(i)}\,\,\,\,\,\widebar{\mathrm{S}}_{\scriptscriptstyle \mathrm{HD}}^{\,\mathrm{low}}
\approx\biggl[\,\frac{A_\mathrm{\scriptscriptstyle 0}}{4 ab}\, \rho \ln^{2}\left(\frac{1}{\rho }\right)\,\biggr]\, \centredComma\,\,\,\,\,\,\,\,\mathrm{and}&\label{eq:PEAsymLow}\\
&\mathrm{(ii)}\,\,\widebar{\mathrm{S}}_{\scriptscriptstyle \mathrm{HD}}^{\,\mathrm{high}}  
\approx\biggl[\,\ln{\rho} + \ln\left( \dfrac{A_\mathrm{\scriptscriptstyle 0}}{ab}\right) +\psi (a) +\psi (b)  - \left(\dfrac{1}{\xi^{2}}\right)\,\biggr]\, \centredComma&\label{eq:AsHighPE}
\end{flalign}
where the superscripts $\mathrm{low}$ and $\mathrm{high}$ correspond to the limits $\rho \to 0$ and $\rho \to \infty$, respectively.
\end{theorem}
\begin{IEEEproof} 
$\textit{Part (i)}$: At high SNR, the threshold scales as $\mu \approx \rho^{-1}$ (cf. Lemma~\ref{eq:lemma_threshold}), while the Meijer-G term vanishes for small arguments, i.e.,
\begin{equation}
\lim_{z \to 0} \,\,\, G_{3,5}^{3,2}\left( z\, \biggr\rvert \, \begin{matrix}
 1, 1, \xi^{2} +1 \\
\xi^{2}, a, b, 0, 0
\end{matrix}\,\right) \,=\,0.
\end{equation}
With these substitutions in Theorem~\ref{eq:thm2}, \eqref{eq:AsHighPE} is proved.

$\textit{Part (ii)}$: The low-order series expansion of the Meijer-G function in~\eqref{eq:ExactPE} for large input argument $z$ (asymp. $\infty$) is
%\fontsize{9.7}{12}\selectfont
\begin{align}
G_{3,5}^{3,2}\left( z \, \biggr\rvert \,\, \begin{matrix}
 1, 1, \xi^{2} +1 \\
\!\xi^{2}, a, b, 0, 0
\end{matrix}\right) \approx \, &\,\dfrac{\sqrt{\pi} \,e^{-2\sqrt{z}}}{ z^{(7-2a-2b)/4}} \,+\,\dfrac{\Gamma ( a) \, \Gamma ( b)}{\xi ^{2}} \,\times \\ &\,\biggl[\,\ln z -\psi (a) -\psi (b)+\dfrac{1}{\xi ^{2}}\,\biggr] \cdot    
\end{align}
%\normalfont
Substituting the above series expansion in~\eqref{eq:ExactPE}, we observe
\begin{align}
\label{eq:CU_firstapprox}
\widebar{\mathrm{S}}_{\scriptscriptstyle \mathrm{HD}} \,&\approx \,\, {\mathcal{K} \,(A_\mathrm{\scriptscriptstyle 0}/ab)}  \left({ab \mu}/{A_\mathrm{\scriptscriptstyle 0}}\right)^{\left({a+b} -\frac{7}{2}\right)\frac{1}{2}} e^{-2\sqrt{{ab \mu}/{A_\mathrm{\scriptscriptstyle 0}}}}.
\end{align}
Comparing~\eqref{eq:CU_firstapprox} with~\eqref{eq:SNRb} derived in connection with the Lemma~\ref{eq:lemma_threshold}, we find that
\begin{equation}
\label{eq:CU_secondapprox}
\widebar{\mathrm{S}}_{\scriptscriptstyle \mathrm{HD}} \,\approx \,\,{\mu \rho}.
\end{equation}
Finally, substituting $\mu$ from ~\eqref{eq:n_below1224} in~\eqref{eq:CU_secondapprox} completes the proof of the low-SNR ASE expansion~\eqref{eq:PEAsymLow} in Theorem~\ref{eq:theorem4}.
\end{IEEEproof}
%In~\cite{ansari2013unified}, with constant beam power allocation (or no power control) at the optical transmitter, the authors have analyzed the ASE performance of this channel under weak or negligible pointing error ($\xi^2 >> 1$) assumption. However, notice that the asymptotic result~\eqref{eq:AsHighPE} is in general valid with no constraint on the amount of pointing error displacement (jitter) present at the optical receiver side. 
%Further, the asymptotic high SNR ASE expression in ~\cite{ansari2013unified} is  devoid of the detrimental role of the $A_\mathrm{\scriptscriptstyle 0}$ parameter on the FSO system performance, although it is incorporated in the pointing error model.
Furthermore, we can deduce a few interesting observations from the derived theorems as follows.
\begin{itemize}
\item By applying $A_\mathrm{\scriptscriptstyle 0} \to 1$ and $\xi^2 \to \infty$, the asymptotic ASEs of the GG turbulence channel `without' pointing errors at low and high SNRs are computed as
\begin{flalign}
&\mathrm{(i)}\,\,\,\,\,\widebar{\mathrm{S}}_{\scriptscriptstyle \mathrm{HD}}^{\,\mathrm{low}}
\approx\biggl[\,\frac{1}{4 ab}\, \rho \ln^{2}\left(\frac{1}{\rho }\right)\,\biggr]\, \centredComma\,\,\,\,\,\,\,\,\mathrm{and}&\label{eq:WPEAsymLow}\\
&\mathrm{(ii)}\,\,\widebar{\mathrm{S}}_{\scriptscriptstyle \mathrm{HD}}^{\,\mathrm{high}}  \approx  \biggl[\,\ln{\rho} + \ln\left(\dfrac{1}{ab}\right) +\psi (a) +\psi (b)\,\biggr]\,\cdot&\label{eq:AsHighNPE}
\end{flalign}
\item The high-SNR ASE degradation solely due to pointing errors, obtained by comparing \eqref{eq:AsHighPE} and~\eqref{eq:AsHighNPE}, is given by
\begin{equation}\label{eq:HighSNRLoss}
\biggl[\dfrac{1}{\xi ^{2}} \,-\, \ln A_\mathrm{\scriptscriptstyle 0}\biggr]\,.
\end{equation}
\item For the GG turbulence channel, the presence of pointing errors decreases the ASE at low SNRs by $A_\mathrm{\scriptscriptstyle 0}$ times.
\end{itemize}
The asymptotic results in~\eqref{eq:PEAsymLow}, \eqref{eq:AsHighPE}, \eqref{eq:WPEAsymLow}, and~\eqref{eq:AsHighNPE} enable comprehensive assessment of turbulence and pointing error effects across diverse terrestrial coherent FSO conditions.

\section{Numerical Results and Discussion}\label{sec:numericalR}
In this section, we present the analytical utility of the derived ASE formulas, supported by extensive numerical results. Previous studies on related topics~\cite{gappmair2011further}, \cite{ansari2013unified}, \cite{lee2016effects}, \cite{doubleGG2015}, and \cite{ansari2015} have suggested, both directly and indirectly, that a wide range of optical turbulence conditions can be achieved by varying the FSO link length. However, link length variations impact the large-scale path loss factor. Additionally, the laser beam waist expands with propagation distance, impacting the pointing error distribution parameters $\xi$ and $A_\mathrm{\scriptscriptstyle 0}$. Notably, these studies did not consider these interdependencies. Understanding the interactions among these channel characteristics is essential, as this work aims to accurately analyze the impact of variations in turbulence and pointing error on the ASE performance of terrestrial FSO systems.

\begin{table}[htbp]
\caption{\textcolor{black}{FSO SYSTEM AND CHANNEL SETTINGS}}\label{tab:FSO_System_settings}
\vspace{-1.0em}
\footnotesize
\centering
\textcolor{black}{
% centering table
\begin{tabular}{@{} l        c         	c @{}}
\toprule\\[-1.35em]
\textbf{\,Parameter}  &  \textbf{\,\,\,\,\,\,Symbol\,\,\,\,\,\,\,\,\,\,\,\,\,\,\,\,\,\,\,\,\,\,\,\,\,\,\,\,\,\,\,\,\,\,\,\,\,\,\,\,\,\,\,} &  \textbf{Value\,\,\,\,} \\[-1.35em] \\ \midrule\\[-1.35em]
\,Optical wavelength&\,\,\,\,\,\,$\lambda_{\mathrm{w}}$\,\,\,\,\,\,\,\,\,\,\,\,\,\,\,\,\,\,\,\,\,\,\,\,\,\,\,\,\,\,\,\,\,\,\,\,\,\,\,\,\,\,& \,\,1550 nm\,\,\,\,\,\\[0.25ex]
\,Tx. Beam waist&\,\,\,\,\,\,$w_{\mathrm{\scriptscriptstyle 0}}$\,\,\,\,\,\,\,\,\,\,\,\,\,\,\,\,\,\,\,\,\,\,\,\,\,\,\,\,\,\,\,\,\,\,\,\,\,\,\,\,\,\,& \,0.45 cm\,\,\,\,\,\\[0.25ex]
\,FSO channel length&\,\,\,\,\,\,$L$\,\,\,\,\,\,\,\,\,\,\,\,\,\,\,\,\,\,\,\,\,\,\,\,\,\,\,\,\,\,\,\,\,\,\,\,\,\,\,\,\,\,&\,300 m\,\,\,\,\,\\[0.25ex]
\,Rx. Aperture radius&\,\,\,\,\,\,$r_{\mathrm{\scriptscriptstyle A}}$\,\,\,\,\,\,\,\,\,\,\,\,\,\,\,\,\,\,\,\,\,\,\,\,\,\,\,\,\,\,\,\,\,\,\,\,\,\,\,\,\,\,& \,1 cm\,\,\,\,\,\\[0.25ex]
\,Tx./Rx. optics eff.&\,\,\,\,\,\,$-$\,\,\,\,\,\,\,\,\,\,\,\,\,\,\,\,\,\,\,\,\,\,\,\,\,\,\,\,\,\,\,\,\,\,\,\,\,\,\,\,\,\,&\,100\% (assumed)\,\,\,\,\,\\[0.25ex]
\,Path loss factor&\,\,\,\,\,\,$-$\,\,\,\,\,\,\,\,\,\,\,\,\,\,\,\,\,\,\,\,\,\,\,\,\,\,\,\,\,\,\,\,\,\,\,\,\,\,\,\,\,\,&\,\,\,Unity (assumed)\,\,\,\,\,\,\,\\[-0.15em]
\bottomrule\\[-1.3em]
\end{tabular}
\\[0.5em]\text{\centering{\textbf{\,Ch. Setting (a): Gamma-Gamma (GG) fading `without' pointing error\phantom{xxxx}\,\,\,\,\,\,\,\,\,\,\,\,\,\,\,}}}\vspace{0.25em}
\centering
% centering table
\begin{tabular}{@{} l c c @{}}
\toprule\\[-1.25em]
\textit{Turbulence\,\,\,\,\,\,\,\,\,\,\,\,\,\,\,\,\,\,\,\,\,\,\,}&\textit{\,\,Rytov\,\,\,\,\,\,\,\,\,\,\,\,\,\,\,}&\textit{\,\,\,\,\,\,GG\,\,\,\,\,\,}\\[-0.2em]
\textit{\,\,\,strength} &\,\,\textit{variance} $\sigma _{{\mathrm{\scriptscriptstyle R}}}^2$\,\,\,\,\,\,\,\,\,\,\,\,\,\,\,&\textit{\,\,\,\,\,\,\,\,distribution parameters\,\,\,}\\[-0.05em] \midrule\\[-1.25em]
\,Weak\,\,\,\,\,\,\,\,\,&\,\,$0.2$\,\,\,\,\,\,\,\,\,\,\,\,\,\,\,&\,\,\,$a \,=\,\, 11.651,\,\,b \,=\, 10.122$\,\,\,\\[0.25ex]
\,Moderate\,\,\,\,\,\,\,\,\,\,&\,\,$1$\,\,\,\,\,\,\,\,\,\,\,\,\,\,\,&\,\,\,$a \,=\,\, 4.3939,\,\,b \,=\, 2.5636$\,\,\,\\[0.25ex]
\,Strong\,\,\,\,\,\,\,\,\,\,&\,\,$4$\,\,\,\,\,\,\,\,\,\,\,\,\,\,\,&\,\,\,$a \,=\,\, 4.3407,\,\,b \,=\, 1.3088$\,\,\,\\[-0.35ex]
\bottomrule\\[-1.3em]
\end{tabular}
\\[0.5em]\text{\centering{\textbf{\,Ch. Setting (b): GG fading with `mild' pointing error ($\sigma_{\mathrm{e}} = 1$ cm)\phantom{x}\,\,\,\,\,\,\,\,\,\,\,\,\,\,\,}}}\vspace{0.25em}
\centering
% centering table
\begin{tabular}{@{} l c c @{}}
\toprule\\[-1.25em]
\textit{Turbulence\,\,\,\,\,\,\,\,\,\,\,\,\,\,\,\,\,\,\,\,\,\,\,}&\textit{\,\,\,Rytov\,\,\,\,\,\,\,\,\,\,\,\,\,\,\,}&\textit{\,\,\,\,\,\,\,Pointing Error\,\,\,\,\,\,}\\[-0.2em]
\textit{\,\,\,strength} &\,\,\,\textit{variance} $\sigma _{{\mathrm{\scriptscriptstyle R}}}^2$\,\,\,\,\,\,\,\,\,\,\,\,\,\,\,&\,\,\,\,\,\,\textit{distribution parameters\,\,\,\,\,}\\[-0.05em] \midrule\\[-1.25em]	
\,Weak\,\,\,\,\,\,\,\,&$0.2$\,\,\,\,\,\,\,\,\,\,\,\,\,\,\,&\,\,$\xi \,=\, 1.8102
,\,A_\mathrm{\scriptscriptstyle 0} \,=\, 0.1526
$\\[0.35ex]
\,Moderate\,\,\,\,\,\,\,\,&$1$\,\,\,\,\,\,\,\,\,\,\,\,\,\,\,&\,\,$\xi \,=\, 2.1977
,\,A_\mathrm{\scriptscriptstyle 0} \,=\, 0.1035
$\\[0.35ex]
\,Strong\,\,\,\,\,\,\,\,&$4$\,\,\,\,\,\,\,\,\,\,\,\,\,\,\,&\,\,$\xi \,=\, 3.5510
,\,A_\mathrm{\scriptscriptstyle 0} \,=\, 0.0397
$\\[-0.35ex]
\bottomrule\\[-1.3em]
\end{tabular}
\\[0.5em]\text{\centering{\textbf{\,Ch. Setting (c): GG fading with `strong' pointing error ($\sigma_{\mathrm{e}} = 4$ cm)\phantom{x}\,\,\,\,\,\,\,\,\,\,\,\,\,\,\,}}}\vspace{0.25em}
\centering
% centering table
\begin{tabular}{@{} l c c @{}}
\toprule\\[-1.25em]
\textit{Turbulence\,\,\,\,\,\,\,\,\,\,\,\,\,\,\,\,\,\,\,\,\,\,\,}&\textit{\,\,\,Rytov\,\,\,\,\,\,\,\,\,\,\,\,\,\,\,}&\textit{\,\,\,\,\,\,\,Pointing Error\,\,\,\,\,\,}\\[-0.2em]
\textit{\,\,\,strength} &\,\,\,\textit{variance} $\sigma _{{\mathrm{\scriptscriptstyle R}}}^2$\,\,\,\,\,\,\,\,\,\,\,\,\,\,\,&\,\,\,\,\,\,\textit{distribution parameters\,\,\,\,\,}\\[-0.05em] \midrule\\[-1.25em]	
\,Weak\,\,\,\,\,\,\,\,&$0.2$\,\,\,\,\,\,\,\,\,\,\,\,\,\,\,&\,\,$\xi \,=\, 0.4525
,\,A_\mathrm{\scriptscriptstyle 0} \,=\, 0.1526
$\\[0.35ex]
\,Moderate\,\,\,\,\,\,\,\,&$1$\,\,\,\,\,\,\,\,\,\,\,\,\,\,\,&\,\,$\xi \,=\, 0.5494
,\,A_\mathrm{\scriptscriptstyle 0} \,=\, 0.1035
$\\[0.35ex]
\,Strong\,\,\,\,\,\,\,\,&$4$\,\,\,\,\,\,\,\,\,\,\,\,\,\,\,&\,\,$\xi \,=\, 0.8877
,\,A_\mathrm{\scriptscriptstyle 0} \,=\, 0.0397
$\\[-0.35ex]
\bottomrule
\end{tabular}}
\end{table}
To ensure a fair comparison, the FSO Tx-Rx system parameters (e.g., optical wavelength, Tx. beam waist, Rx. aperture) are kept fixed. In addition, a horizontal terrestrial FSO link of `fixed length' is considered, ensuring that the path loss factor remains unchanged. \textcolor{black}{A horizontal distance of 300 m, representative of short-to-medium haul FSO links, is selected to capture realistic turbulence and pointing error effects while keeping other impairments negligible (see Remark~\ref{eq:rem1}). To explicitly account for the acquisition, tracking, and pointing (ATP) subsystem, three representative GG channel settings are considered:
(a) \emph{No pointing error}, corresponding to an ideal or “genie-aided’’ ATP subsystem that fully eliminates jitter, serving as a turbulence-limited benchmark;
(b) \emph{Mild pointing error}, corresponding to a realistic, well-designed ATP subsystem with small residual jitter; and
(c) \emph{Strong pointing error}, corresponding to a poorly designed or absent ATP subsystem (i.e., no ATP), where jitter dominates. Although such severe jitter levels exceed typical values observed in practical FSO systems employing ATP, they are nevertheless included for theoretical completeness, as they highlight the transition 
from turbulence-limited to jitter-dominated operation.}

\textcolor{black}{These FSO link settings cover the practical range of ATP capabilities and provide a unified framework for analyzing turbulence-limited, jitter-limited, and intermediate regimes.}

To span a realistic weak-to-strong turbulence range while maintaining constant path loss, we recall that the turbulence strength is governed by the Rytov variance
\begin{align}
\sigma _{{\mathrm{\scriptscriptstyle R}}}^{2} = 1.23 C_{\mathrm{n}}^2 k_{\mathrm{w}}^{7/6}{L^{11/6}},
\end{align}
which underscores its dependence on the index-of-refraction structure parameter $C_{\mathrm{n}}^2$, as discussed below.
\begin{remark}
For a near-ground horizontal propagation path, the index-of-refraction structure parameter $C_{\mathrm{n}}^{2}$ remains unchanged and varies with height above ground~\cite[Section~12.2]{andrews2005laser}. A wide range of optical turbulence conditions is achievable at different altitudes through $C_{\mathrm{n}}^{2}$; being stronger near the ground and weaker at higher elevations~\cite[Ch.~2]{beland1993}.
\end{remark}

\begin{figure}[!t]
\centering
\includegraphics[height=0.775\linewidth]{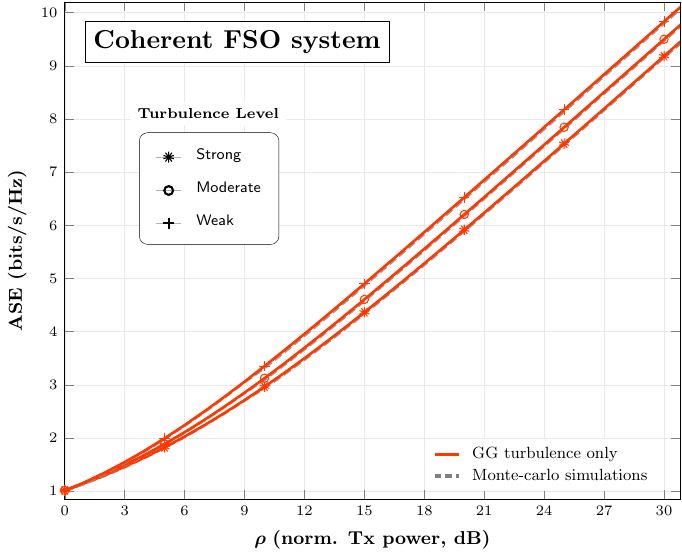}
\caption{Average spectral efficiency achievable over a GG turbulence channel \emph{without} pointing errors under adaptive coherent (I/Q) FSO transmission and synchronous heterodyne detection at the receiver. The average received SNR is the same as $\rho$  since the mean channel gain $\mathbb{E} [\lambda] \,=\, 1$.}
\label{fig:ExWPE}
%\vspace{1.0em}
\end{figure}

\begin{figure}[!t]
\centering
\includegraphics[height=0.775\linewidth]{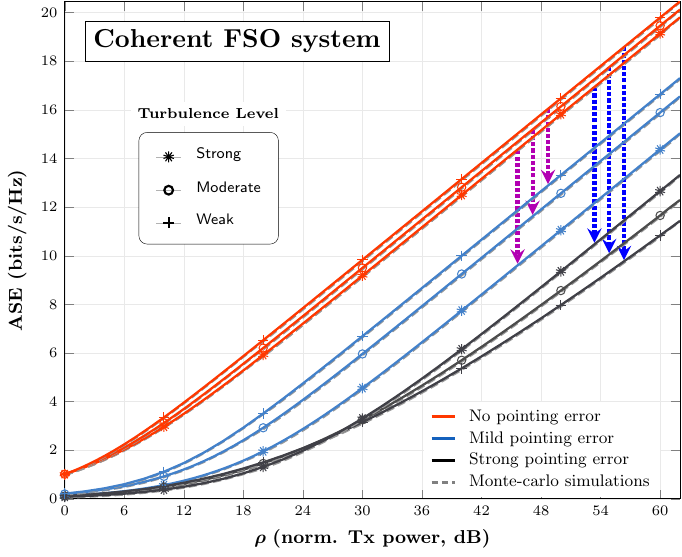}
\caption{Average spectral efficiency achievable over a GG turbulence channel \emph{with} pointing jitter under adaptive coherent FSO transmission and synchronous heterodyne detection at the receiver. The received SNR is less than $\rho$ by a factor equal to the mean channel gain $\mathbb{E} [\lambda]$.}
%\vspace{1em}
\label{fig:PE3}
\end{figure}

\begin{figure}[!t]
\centering
\includegraphics[height=0.76\linewidth]{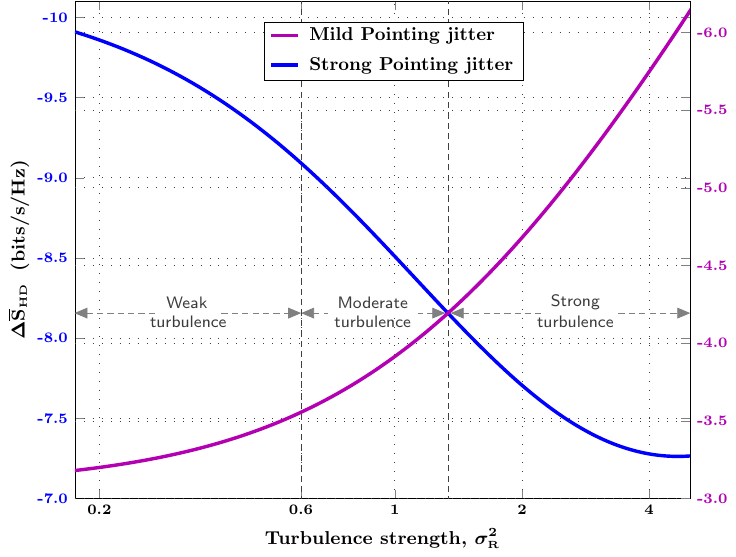}

\caption{ASE degradation at high SNRs versus turbulence strength in a GG turbulence channel, for mild and strong pointing jitter, under adaptive  coherent FSO transmission and synchronous heterodyne detection at the receiver.}
\label{fig:PE6}
\end{figure}

\begin{figure}[!b]
\centering
\includegraphics[height=0.775\linewidth]{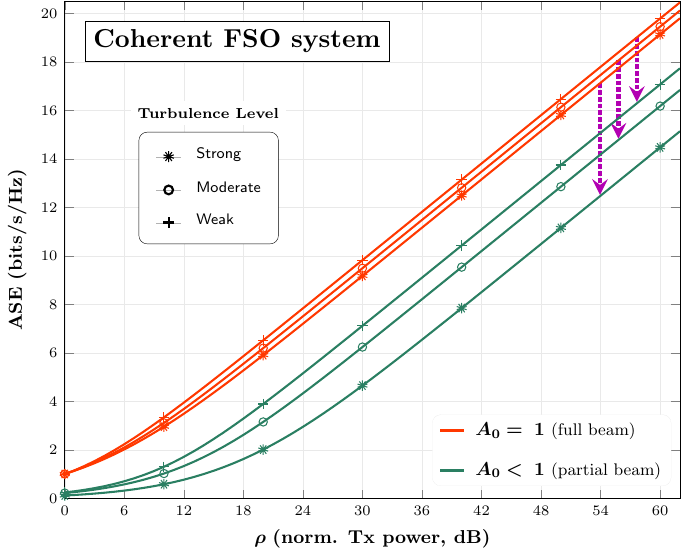}
\caption{\textcolor{black}{Average spectral efficiency over a GG turbulence channel without pointing errors under adaptive coherent FSO transmission and synchronous heterodyne detection: \textcolor{black}{$A_\mathrm{\scriptscriptstyle 0} = 1$ corresponds to channel condition with full-beam capture, whereas the values of $A_\mathrm{\scriptscriptstyle 0}$ listed in Table~\ref{tab:FSO_System_settings} represent channel conditions with partial-beam capture. The definition of $A_\mathrm{\scriptscriptstyle 0}$ is given in~\eqref{eq:redefined_A0}.}}}
\label{fig:PE30}
\end{figure}

\begin{figure}[t]
\centering
\includegraphics[height=0.775\linewidth]{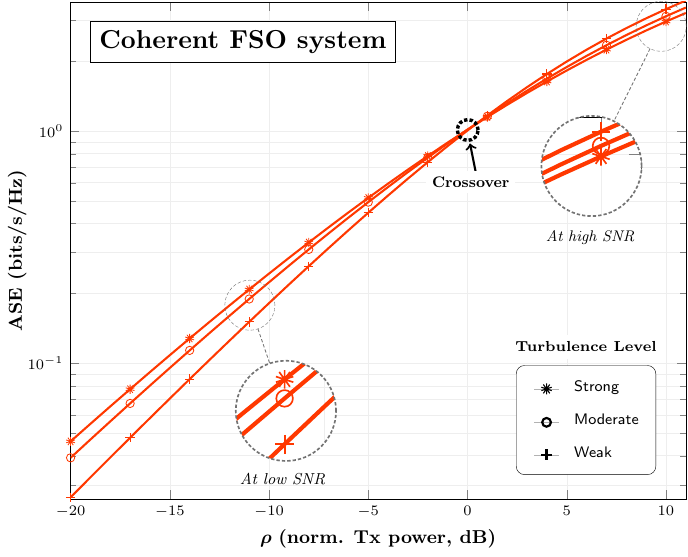}
\caption{\textcolor{black}{Average spectral efficiency achievable at low SNRs for a GG turbulence channel \emph{without} pointing jitter under adaptive coherent FSO transmission and synchronous heterodyne detection at the receiver (see Table~\ref{tab:FSO_System_settings} for the turbulence parameters).}}
\label{fig:WPEAsymLowHD}
\end{figure}

\begin{figure}[!b]
\includegraphics[height=0.775\linewidth]{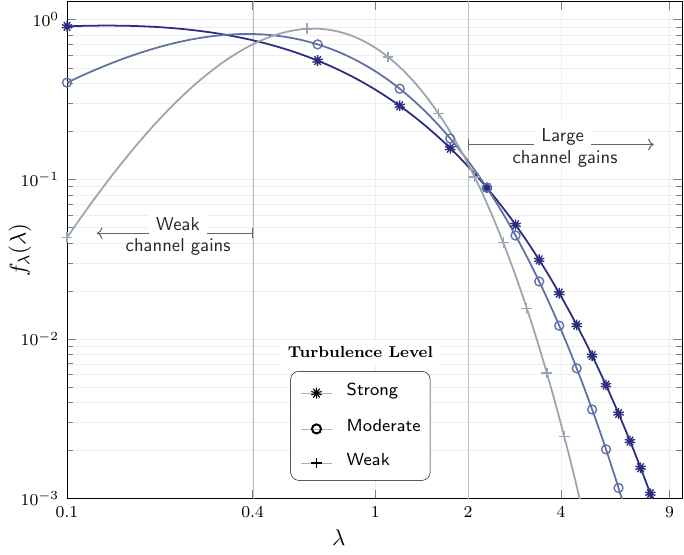}
\caption{\textcolor{black}{Probability density function (PDF) of the GG channel gain for the weak, moderate, and strong turbulence profiles described in Table~\ref{tab:FSO_System_settings}. Notice the increasing probability mass for higher channel gains with increasing turbulence strength.
}}
\label{fig:highergainsPDF}
\end{figure}

\begin{figure}[!t]
\centering
\includegraphics[scale=0.73]{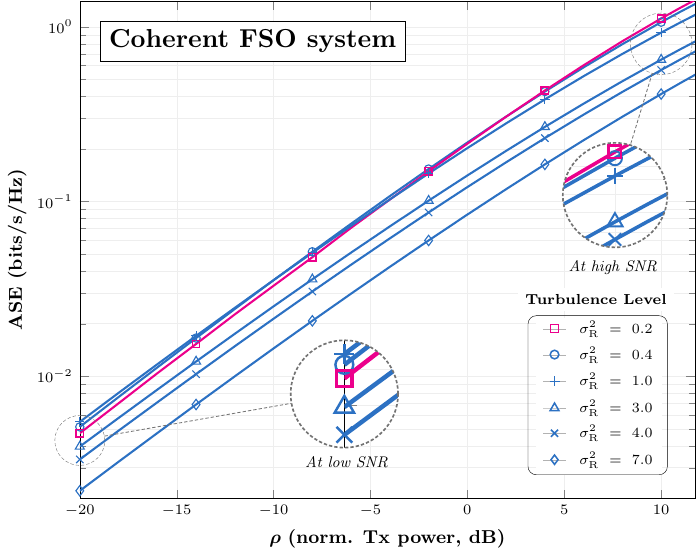}
\caption{\textcolor{black}{Average spectral efficiency achievable in the low-SNR regime over a GG turbulence channel with \emph{mild} pointing jitter under adaptive coherent FSO transmission and synchronous heterodyne detection at the receiver.}}
\label{fig:PEAsymLowHD2}
\end{figure}

\begin{figure}[!t]
\centering
\includegraphics[height=0.775\linewidth]{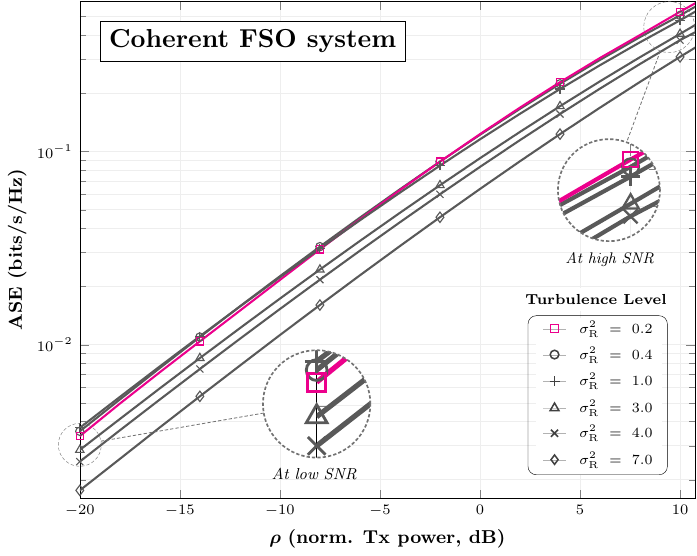}
\caption{\textcolor{black}{Average spectral efficiency achievable in the low-SNR regime over a GG turbulence channel with \emph{strong} pointing jitter under adaptive coherent FSO transmission and synchronous heterodyne detection at the receiver.}}
\label{fig:PEAsymLowHD1}
\end{figure}

A few additional important considerations to be noted before proceeding with the numerical results.
\begin{itemize}
\item The horizontal axis in Figures~\ref{fig:ExWPE}$-$\ref{fig:PEAsymLowHD1} represents $\rho$ (in dB scale), i.e., the normalized transmit power, rather than the received SNR, ensuring a fair and consistent comparison. Recall from~\eqref{eq:scaledPower} that $\rho := P_{\mathrm{avg}}/\sigma_{\mathrm{\scriptscriptstyle OLO}}^{2}$ is the noise-normalized transmit power budget in all the derived ASE formulae. If required, the average received SNR can be computed by scaling $\rho$ with the mean channel gain $\mathbb{E}[\lambda]$.  
\item All ASE values (bits/sec/Hz is the unit of measurement) are plotted by scaling the derived expressions by ${1}/{(\ln 2)}$.
\end{itemize} 

\noindent
\emph{1) Exact ASE behavior:} Based on Theorem~\ref{eq:thm2}, the exact ASE for HD-detection based coherent optical channel under GG turbulence, with and without pointing errors, is computed and shown in Figures~\ref{fig:ExWPE} and~\ref{fig:PE3}, respectively. The analytical results closely match Monte Carlo simulation outcomes across a wide SNR range and under various atmospheric turbulence regimes and pointing jitter conditions.

\noindent
\emph{Discussion:} The following key observations are drawn from Figures~\ref{fig:ExWPE} and~\ref{fig:PE3}:
\begin{itemize}
\item ASE deteriorates with increasing turbulence at high SNR, as shown in Fig. \ref{fig:ExWPE}; the PDF concentrates more heavily at lower channel gains as turbulence increases (cf. Fig.~\ref{fig:highergainsPDF}). 
\item \textcolor{black}{As shown in Fig.~\ref{fig:PE3}, ASE drops significantly at high SNRs in the presence of pointing errors. This is due to the effective SNR reduction at the receiver by the factor $\mathbb{E}[\lambda]$: milder jitter levels result in a moderate reduction in effective SNR, while  severe jitter leads to a significant degradation in system performance.}

\item Interestingly, Fig.~\ref{fig:PE3} reveals that, under strong pointing errors, ASE improves with turbulence in the high-SNR regime: \textcolor{black}{the transition from ASE loss to gain initiates at norm. tx. power $\rho = 30$ dB, and  correspondingly to a received SNR of $(30 - 17.58) \approx 12.42$ dB (since $\mathbb{E}[\lambda] = -17.58$ dB).} To explain this counterintuitive behavior, an asymptotic high-SNR analysis is carried out next.
%Interestingly, Fig.~\ref{fig:PE3} shows that spectral efficiency improves with higher turbulence at high SNRs for a strong pointing error channel setting. \textcolor{black}{Notice that the transition (shift from ASE deterioration to improvement)
%begins at $30$ dB of transmit SNR which translates to $(30 - 17.58) \approx 12.42$ dB of received SNR (justified in the caption of Fig.~\ref{fig:PE3}).} To understand this seemingly counterintuitive behavior, we proceed with an asymptotic analysis in the high-SNR regime.
\end{itemize}

\noindent
\emph{2) ASE behavior at High SNRs:} The high-SNR asymptotic ASE behavior in~\eqref{eq:AsHighPE}, as established in Theorem~\ref{eq:theorem4}, invites comparison with the ASE of an \emph{ideal AWGN} channel, which is asymptotically given by $\ln  \rho$. The resulting difference---being strictly negative---quantifies the \emph{spectral efficiency degradation} at high SNRs due to the combined GG turbulence and pointing jitter impairments, and is given by
\begin{align}\label{eq:penaltyPE_high}
\Delta\overline{\mathrm{S}}_{\mathrm{\scriptscriptstyle HD}} \,\triangleq\,\ln \left(\frac{A_\mathrm{\scriptscriptstyle 0}}{ab}\right) \,+\,\psi (a) \,+\,\psi (b) \,-\, \left(\frac{1}{\xi^{2}}\right)\cdot
\end{align}
\noindent
\emph{Discussion:} Fig.~\ref{fig:PE6} depicts the spectral efficiency penalty \eqref{eq:penaltyPE_high} plotted as a function of atmospheric turbulence strength for the mild and strong pointer error channel settings in Table~\ref{tab:FSO_System_settings}. Recall that a fixed-length horizontal FSO link is considered, with a laser beam of waist $w_\mathrm{\scriptscriptstyle 0} = 4.5$ mm at the transmitter (see Table~\ref{tab:FSO_System_settings}). The beam footprint at the receiver is broadened due to beam divergence caused by turbulence over the link length. For the strong pointing error channel setting, the enlarged beam footprint leads to
\begin{itemize}
\item a reduction in collected power at the fixed-aperture receiver (i.e., decreased $A_\mathrm{\scriptscriptstyle 0}$); and
\item a possible mitigation of the pointing jitter effect (i.e., increased $\xi$).
\end{itemize}

%As shown in Fig.~\ref{fig:PE6}, for the GG turbulence channel under strong pointing error setting, the spectral efficiency at high SNR improves with turbulence in the realistic turbulence range $0 < \sigma_{\mathrm{R}}^2 \leq 4$ since the reduction in jitter impact outweighs the power loss. %However, for $\sigma_{\mathrm{R}}^2 \,>\, 6$, pointing jitter becomes negligible while the continued drop in $A_\mathrm{\scriptscriptstyle 0}$ dominates, resulting in a net loss in spectral efficiency. 
%The observed ASE varies significantly with turbulence, ranging from approximately $-10$ to $-7.1$ bits/s/Hz. It is remarkable that an ASE improvement of almost $3$ bits/s/Hz is achievable with turbulence at high SNRs in the presence of strong pointing errors.
%
%On the other hand, for the case of GG fading with mild pointing error setting, no such trade-off between decreasing $A_\mathrm{\scriptscriptstyle 0}$ and increasing $\xi$ happens. This is simply due to the fact that the $\xi$ is already in a good condition and does not improve much with turbulence while $A_\mathrm{\scriptscriptstyle 0}$ decays significantly. This overall leading to ASE deterioration with turbulence. This high-SNR spectral efficiency penalty is 
%%\begin{align}\label{eq:penalty_high}
%%\widebar{S}_{\textrm{penalty}} = \psi(a) + \psi(b) - \ln(ab),
%%\end{align}
%shown as the blue curve in Fig.~\ref{fig:PE6}. This penalty remains relatively significant$-$ranging from approximately $-6.1$ to $-3.2$ bits/s/Hz. Similar arguments hold for ASE deterioration with turbulence for the GG fading without pointing error 
%case.

\noindent
\textcolor{black}{As shown in Fig.~\ref{fig:PE6}, for the GG fading with `strong' pointing errors, the ASE at high SNR increases with turbulence over the practical range $0 < \sigma_\mathrm{\scriptscriptstyle R}^2 \leq 4$. This atypical behavior arises because the reduction
in the pointing jitter effect more than compensates for the loss in
collected power at the receiver. Consequently, the ASE penalty decreases approximately from $-10$ to $-7.2$ bits/s/Hz, yielding a gain of nearly $2.8$ bits/s/Hz in the high SNR regime.}

\textcolor{black}{For the GG fading with `mild' pointing errors setting, no such trade-off is observed, as increased $\xi$ induces marginal improvement while $A_\mathrm{\scriptscriptstyle 0}$ continues to decay significantly, leading overall to a monotonic ASE loss with turbulence. The corresponding high-SNR penalty, illustrated by the purple curve in Fig.~\ref{fig:PE6}, ranges from $-3.2$ to $-6.1$ bits/s/Hz.}

\textcolor{black}{In the GG fading without pointing errors, on the other hand, the penalty at high SNRs is described by (see~\eqref{eq:AsHighNPE})
\begin{align}\label{eq:penalty_high}
\Delta\overline{\mathrm{S}}_{\mathrm{\scriptscriptstyle HD}} \,=\,  \psi (a) \,+\,\psi (b) \,-\, \ln\left(ab\right),
\end{align}
which results in a small ASE degradation, as can be inferred from Fig.~\ref{fig:PE3}  \textcolor{black}{(Fig.~\ref{fig:PE30} provides a more refined characterization)}.}

%\begin{remark}
%The penalty curve (red) in Fig.~\ref{fig:PE6} is valid only for the considered FSO system and channel settings described in Table~\ref{tab:FSO_System_settings}. Keeping the FSO system parameters the same, a similar penalty effect follows for other FSO link lengths; more penalty for shorter link lengths in weak turbulence conditions and vice-versa.
%\end{remark}

Two additional observations can be drawn from the ASE results  as follows.
\begin{itemize}
\item \textcolor{black}{A common inconsistency in the literature is the omission of the parameter $A_\mathrm{\scriptscriptstyle 0}$ when modeling optical fading without pointing errors. $A_\mathrm{\scriptscriptstyle 0} = 1$ suggests full beam capture which is completely justifiable for short-haul optical links where beam divergence is small and receive aperture is wide enough to capture the whole beam. $A_\mathrm{\scriptscriptstyle 0} < 1$ implies partial capture due to strong beam divergence and finite aperture effects. Both cases are realistic in practice, but neglecting $A_\mathrm{\scriptscriptstyle 0} < 1$, when not physically justified, can lead to an overestimated performance, as illustrated in Fig.~\ref{fig:PE30}.}

\textcolor{black}{From Fig.~\ref{fig:PE30}, the degradation in ASE due to varying turbulence (weak to strong) under the $A_\mathrm{\scriptscriptstyle 0} = 1$ condition is modest at only $0.5$ bits/s/Hz. In contrast, when $A_\mathrm{\scriptscriptstyle 0} < 1$, the ASE degradation due to turbulence increases significantly to 2.6 bits/s/Hz. This contrast highlights the importance of modeling $A_\mathrm{\scriptscriptstyle 0}$ accurately. The vertical downward-pointing arrows in Fig.~\ref{fig:PE30} further indicate that the ASE reduction associated with $A_\mathrm{\scriptscriptstyle 0} < 1$ is more severe under strong turbulence and less pronounced under weak turbulence.}

\item For the \textit{fixed-length} optical channel with a constant FSO system configuration, as considered here, an important observation arises regarding the impact of increasing pointing jitter. Recall that $\sigma_{\mathrm{e}}$ denotes the pointing-jitter standard deviation. At high SNR, the resulting ASE degradation (in nats/s/Hz) attributable solely to jitter can be obtained directly from~\eqref{eq:AsHighPE} and is expressed as
\begin{align}
\Delta\widebar{\mathrm{S}}_{\scriptscriptstyle \mathrm{HD}}^{\xi}  \,&\approx\,  \left[\,\frac{1}{\xi^2_{\mathrm{new}}} - \frac{1}{\xi^2_{\mathrm{old}}}\,\right] \,=\, \frac{4 (\sigma_{\mathrm{e,\,new}}^2 -\sigma_{\mathrm{e,\,old}}^2)}{w_{{\mathrm{\scriptscriptstyle L_{eq}}}}^2}\,\centredComma
\end{align}
where $w_{{\mathrm{\scriptscriptstyle L_{eq}}}}$ denotes the received equivalent beam waist (see just above Eq.~\eqref{eq:fractionPowA0}). As turbulence intensity increases, $w_{\mathrm{L_{eq}}}$ also increases, and vice versa. Hence, the ASE loss associated with a given increment in pointing jitter exhibits an inverse dependence on turbulence strength: the penalty is reduced under stronger turbulence fluctuations and amplified under weaker fluctuations, as illustrated by the vertical down-arrowed lines in Fig.~\ref{fig:PE3}.
\end{itemize}

\begin{figure}[!t]
\centering
\includegraphics[height=0.775\linewidth]{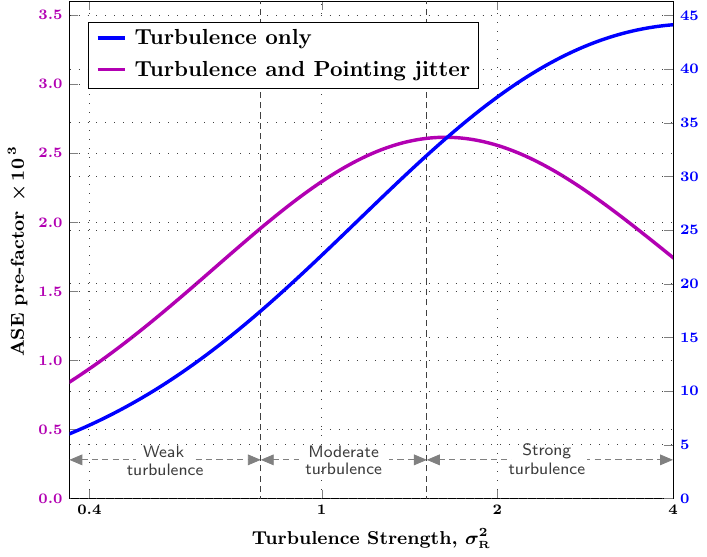}
\caption{\textcolor{black}{Variation of the low-SNR ASE pre-factor for the GG turbulence channel \emph{with} and \emph{without} pointing jitter under adaptive  coherent FSO transmission and synchronous heterodyne detection at the receiver.}}\label{fig:PE8}
\end{figure}

\vspace{-1.25em}
\textcolor{black}{\begin{remark}\label{eq:rem3}
The Figures~\ref{fig:WPEAsymLowHD}--\ref{fig:PE8} are grouped according to the previously discussed cases $A_\mathrm{\scriptscriptstyle 0}=1$ and $A_\mathrm{\scriptscriptstyle 0}<1$.
Figures~\ref{fig:WPEAsymLowHD} and \ref{fig:highergainsPDF} correspond to turbulence-only conditions with $A_\mathrm{\scriptscriptstyle 0}=1$, whereas Figs.~\ref{fig:PEAsymLowHD2} and \ref{fig:PEAsymLowHD1} consider composite turbulence and pointing-error conditions with the corresponding $A_\mathrm{\scriptscriptstyle 0}<1$ values listed in Table~\ref{tab:FSO_System_settings}. In Fig.~\ref{fig:PE8}, both channel cases are analyzed: the turbulence-only condition is characterized by $A_\mathrm{\scriptscriptstyle 0}=1$, while the turbulence with pointing errors is characterized by $A_\mathrm{\scriptscriptstyle 0}<1$.
\end{remark}}

\noindent
\emph{3) Exact and asymptotic ASE behavior at Low SNRs:} As highlighted in the introduction, understanding the ASE limit of terrestrial FSO links at low SNRs is of practical interest. At low power budgets, power control of the transmitted laser beam becomes critical, requiring efficient exploitation of channel fading.

\noindent
\emph{Discussion:} Fig.~\ref{fig:WPEAsymLowHD} presents the exact ASE for the GG fading channel without pointing error under the HD detection scheme at low SNRs. Somewhat surprisingly, the low-SNR spectral efficiency improves with increasing turbulence. This agrees with~\eqref{eq:WPEAsymLow}, which shows that strengthening GG turbulence from profile $(a_i, b_i)$ to $(a_j, b_j)$ yields the ASE improvement at sufficiently low SNRs, determined by the ratio of the corresponding pre-factors, which can be expressed as
\begin{align}\label{eq:improvement_WPE_high}
\text{ASE pre-factor ratio} \,\approx\, \frac{a_j b_j}{a_i b_i}\,\cdot
\end{align}
\noindent
This ASE improvement with turbulence at low SNRs is attributed to the enhanced distribution of higher channel gains with increasing turbulence, as shown in Fig.~\ref{fig:highergainsPDF}. At low SNR, the transmitter, assuming full CSI, adapts the beam power optimally, allocating more power to stronger channel gains while avoiding power wastage on weaker states. Consequently, power adaptation yields a net ASE gain at low SNRs, as confirmed by the numerical results in Fig.~\ref{fig:WPEAsymLowHD}. For instance, at a target SNR of $-10$ dB/Hz, the ASE improves from $0.1816$ bits/s/Hz under weak turbulence to $0.2434$ bits/s/Hz under strong turbulence ($34\%$ increase). In a wavelength-division multiplexed (WDM) system with $50$ GHz optical bandwidth per channel, this corresponds to an increase from about $9.08$ Gbps to $12.172$ Gbps per channel for the same transmit power budget. Such per-channel low-SNR conditions are not typical in current deployments but may plausibly arise in future ultra-wideband WDM coherent FSO systems where a fixed eye-safety-limited transmit power must be shared among many carriers, with atmospheric variability (e.g., haze or fog) further accentuating this effect. At a lower target SNR of $-15$ dB/Hz, the ASE increases from about $3.52$ Gbps/channel to $5.42$ Gbps/channel ($53\%$ gain). Moreover, the channel’s ASE per unit SNR$-$known as the wideband slope~\cite{verdu2002}$-$improves as the SNR decreases: under strong turbulence it rises from $2.43$ units at $-10$ dB to $3.43$ units at $-15$ dB ($41\%$ increase), while under weak turbulence it increases from $1.82$ to $2.27$ units ($25\%$ gain). Here, `unit' denotes bits/s/Hz per unit SNR, with SNR expressed in linear ratio.

In contrast, the effect of turbulence on the spectral efficiency of the optical channel with pointing error at low SNRs is more involved. As depicted in Figures~\ref{fig:PEAsymLowHD2}--\ref{fig:PEAsymLowHD1}, the spectral efficiency initially shows a slight improvement in the weak-to-moderate turbulence regime, but then begins to degrade as turbulence strength increases from moderate to strong levels.

\noindent
This unusual behavior can be explained using the asymptotic low-SNR spectral efficiency result in~\eqref{eq:PEAsymLow} from Theorem~\ref{eq:theorem4}. The net impact of atmospheric turbulence and pointing error on the optical channel's ASE at low SNR is captured by the scaling $A_\mathrm{\scriptscriptstyle 0}/(4ab)$, which we rewrite as
\begin{align}\label{eq:lowSNRscale}
\textrm{ASE pre-factor} \,\triangleq\,  A_\mathrm{\scriptscriptstyle 0} \times \dfrac{1}{4ab}.
\end{align}

\noindent
Fig.~\ref{fig:PE8} illustrates the variation of the pre-factor in~\eqref{eq:lowSNRscale} (purple curve) valid for turbulence with pointing error, alongside the factor ${1}/{4ab}$ (blue curve), which applies to pure optical turbulence (see~\eqref{eq:WPEAsymLow}). A comparison of the two curves reveals that the parameter $A_\mathrm{\scriptscriptstyle 0}$, representing the fraction of collected power, decreases with turbulence$-$initially slowly in the weak fluctuation regime and then more rapidly under strong turbulence, while the factor ${1}/{(4ab)}$ increases. For further insight into how $A_\mathrm{\scriptscriptstyle 0}$ varies with turbulence, refer to the $A_\mathrm{\scriptscriptstyle 0}$ values in Table~\ref{tab:FSO_System_settings}, which provide typical values for weak, moderate, and strong turbulence conditions. In general, except under weak fluctuations, the channel's spectral efficiency under pointing errors at low SNR degrades with increasing turbulence over a broad range of moderate-to-strong conditions.

Interestingly, the asymptotic result~\eqref{eq:PEAsymLow} indicates that the influence of pointing jitter strength $\xi^2$ on the adaptive coherent FSO channel's ASE vanishes in the low-SNR regime: the transmitter operates only during channel peaks, and it can be verified that the distribution of these peaks is largely insensitive to $\xi^2$. In \cite{tall2012}, the low-SNR ASE analysis for mobile RF channels reveals two key insights: power adaptation approaches an asymptotically optimal on--off strategy, and transmitter CSI reduces to one-bit feedback. These structural simplifications are expected to hold for terrestrial FSO channels at low SNR, with practical relevance to coherent FSO system design.

\section{Conclusion}\label{sec:conclude}
In this work, we have analyzed the exact average spectral efficiency performance of coherent terrestrial FSO communications over gamma-gamma turbulence with optimal transmitter adaptation. The analysis also highlighted the detrimental impact on the performance due to pointing error impairments typical in terrestrial FSO links.

The proposed exact spectral efficiency solutions capture the impact of fading and pointing error parameters, with particular emphasis on the high and low SNR regimes, and highlight an interplay between $A_\mathrm{\scriptscriptstyle 0}$ and $\xi^2$ as turbulence and pointing error conditions vary. \textcolor{black}{Under mild or negligible pointing error condition$-$typical of practical systems with well-designed ATP$-$the expected degradation of spectral efficiency with increasing turbulence is observed. In contrast, for channels with strong pointing error condition$-$typical of systems with poorly designed or absent ATP, which may occur in low-cost deployments$-$a notable improvement in spectral efficiency is observed at high SNRs as turbulence increases from weak conditions, since the reduction in pointing jitter outweighs the loss in received power. Overall, these results provide insights across the full spectrum of ATP capabilities, from practical well-designed systems to theoretical or low-cost configurations with no or poorly performing ATP.}

At low SNRs, we have shown that the average spectral efficiency can improve with turbulence due to two main factors: i) the distribution of higher fading gains improves with turbulence, and ii) the transmitter efficiently exploits these higher gains through beam power control. However, in the presence of pointing error, the distribution of higher gains deteriorates significantly with turbulence, resulting in a loss of spectral efficiency at low SNRs. \textcolor{black}{From a system design perspective, operation at low SNRs is particularly relevant for terrestrial coherent FSO links for two reasons: first, the stringent per-hertz optical power limitations plausible in future ultra-wideband, long-haul deployments, and second, the superior energy efficiency of wideband communication systems when operated in the power-limited regime. In this light, the spectral efficiency characterization developed in this work provides timely and practically significant insights into the design of next-generation coherent FSO systems.}

\textcolor{black}{As future work, the proposed spectral-efficiency framework can be extended to multi-aperture (MIMO) systems, which we identify as a promising direction for advancing research on next-generation terrestrial coherent FSO systems. Similarly, leveraging the derived ASE limits as a reference for analyzing and optimizing practical coherent FSO implementations represents another valuable avenue for further investigation.}

%\textcolor{black}{As future work, tThe generalization of spectral-efficiency characterization for long-haul terrestrial FSO channels is a line of work that can be pursued further. In particular, future studies should investigate the impact of practical channel impairments$-$such as boresight misalignment, beam wander, and angle-of-arrival fluctuations$-$which become especially pronounced in long- and medium-haul links under strong turbulence conditions.}
\balance

\bibliographystyle{IEEEtran}

\bibliography{references}

\end{document}